\documentclass[prl,twocolumn,showpacs,preprintnumbers,amsmath,amssymb,revtex4-1]{revtex4-1}
\usepackage{amsfonts, amsmath, bm, bbm, graphicx, epsfig, epstopdf}
\usepackage{dcolumn}
\usepackage{textcomp}
 \usepackage[caption=false]{subfig}

\DeclareMathOperator{\sech}{sech}

\begin{document}
\title{Rydberg Noisy-Dressing \\ and applications in making soliton-molecules and droplet quasi-crystals}

\author{Mohammadsadegh Khazali}
\affiliation{ Department of Physics, Sharif University of Technology, Tehran 14588, Iran \\
 School of Nano Science, Institute for Research in Fundamental Sciences (IPM), Tehran 19395-5531, Iran\\
 Department of Physics and Astronomy, Aarhus University , DK 8000 Aarhus C, Denmark}

\date{\today}
\begin{abstract}
The current advances in the field of ultra-cold atoms and atomic traps recall new controllable long-range interactions. These interactions are expected to extend the range of realizable quantum algorithms as well as providing new control mechanisms for the new types of quantum matters. This article presents special inter-atomic interactions between Rydberg-dressed atoms by manipulating lasers' line-width. The new interaction features a hybrid spatial profile containing plateaus and Gaussian peaks. Combined with dynamic individual control over the sign and strength of these interaction elements, Rydberg noisy-dressing (RnD) scheme provides a valuable interaction toolbox for quantum technology. As an example, RnD's application in making  stable gigantic 3D soliton molecules and in the formation of quasi-periodic droplet-crystals are discussed. 
 \end{abstract}

\maketitle

Strong long-range interaction of highly excited atoms lead to a wide range of applications in Quantum technology and science \cite{Jak00,Luk01,Saf10,Ada19,Kha19,Kha20,Fri05,Gor11,Tia19,And16,Par15,Kha17}.  However, their radiative decay has prevented the observation of long-time excitation dynamics or coherent interaction effects on atomic motion. Furthermore, the diverging spatial scaling of Rydberg interaction ($\propto r^{-3}$ or $r^{-6}$) was limiting the range of applications.  
Lightly dressing atoms with Rydberg states \cite{ Bou02,Hen10, Pup10, Hon10,Joh10,Gau16,Zei16} promises enhanced lifetimes while providing sizeable interactions. 
More interesting, Rydberg dressing results to a plateau type interaction, where the atoms within
the blockade volume would experience a Kerr-type interaction \cite{Kha16,Kha18}. This type of interaction  has led to interesting applications in  macroscopic entanglement generation and frequency metrology \cite{Nor19,Bou02,Gil14,Kha16,Kea16,Kha18}, many-body physics in quantum gases \cite{Cin10,Mau11,Hen12,Dau12,Cin14,XLi15}, synthetic quantum magnets \cite{Gla14,Gla15,Bij15} and in quantum computation \cite{ Kea13,Kea15,Mob13}.

By adding laser line-width to the dynamics of in-resonance Rydberg dressing scheme \cite{Gau16}, this article shows that laser noise could be used as a controlling knob for manipulating the interaction features. 
The new Rydberg noisy-dressing (RnD) interaction features hybrid interaction profile containing  soft-core, barriers and wells. Having dynamical control over sign, strength and distance of these elements provides a valuable toolbox for the implementation of quantum technology.
Shell type interaction provided with RnD, could be used for smart addressing of desired interacting sites in atomic traps \cite{Nog14,Xia15,Zei16,Coo18,Nor18,Hol19,Sas19,Wan16,Bar18}, ideal for the implementation of quantum annealing \cite{Gla17},  Quantum simulation \cite{Wei10} and spin-ice formation\cite{Gla14}.
RnD of Bose-Einstein condensate (BEC) opens new research perspectives where this article discusses  its application in making the droplet quasi-crystals as well as making stable giant soliton-molecules in details.

 \begin{figure}[!h]
\centering
    \scalebox{0.47}{\includegraphics*[viewport=110 230 630 450]{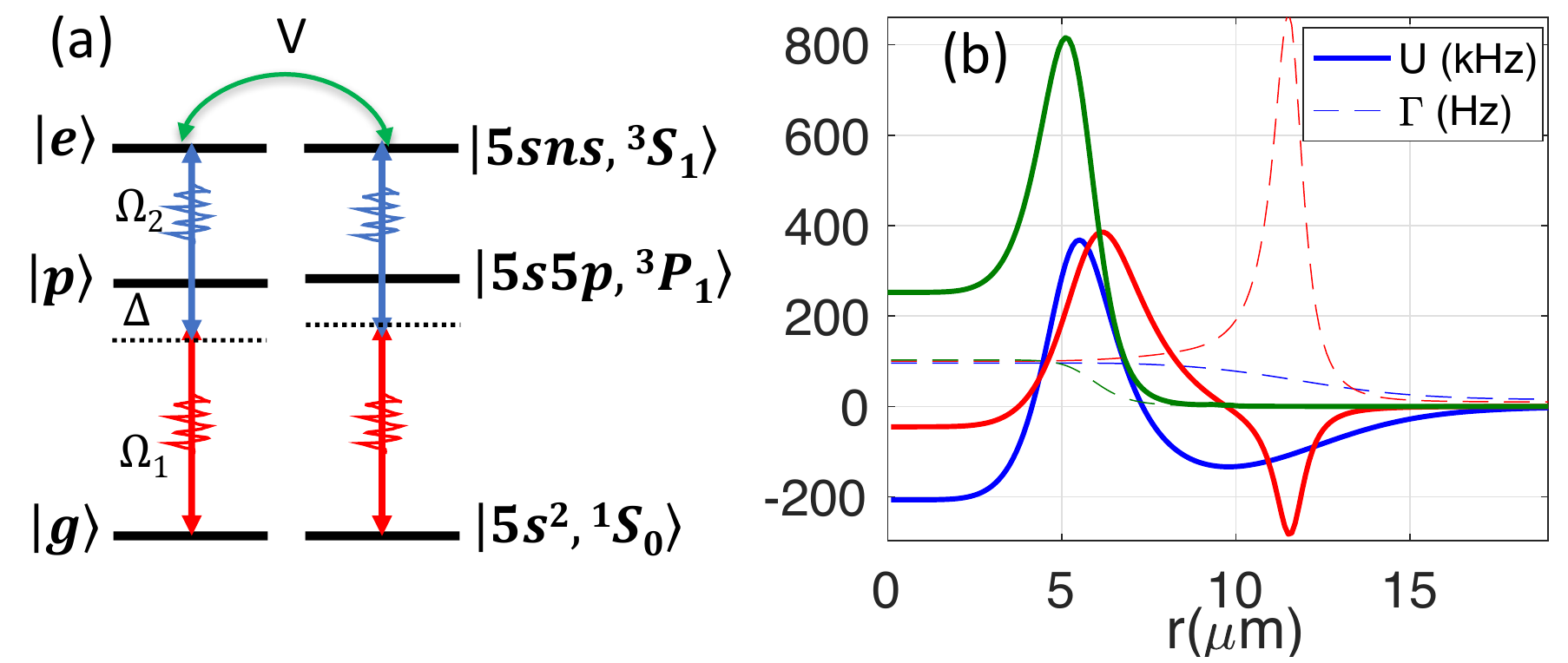}} 
\caption{Rydberg noisy dressing (RnD) interaction. (a) Level scheme of two-photon in-resonance dressing of Sr atoms under EIT condition with noisy lasers.
(b) Sample spatial profiles of RnD interaction (in kHz) containing soft-core, barrier and well elements. These  elements could be controlled individually by manipulating laser parameters. Loss rate per atom  $\Gamma$ is plotted by dashed lines (in Hz). These interactions are used in making soliton-molecules and droplet quasi-crystals in this article. 
  Laser  parameters are explained in Fig. \ref{MolFusibility},\ref{QuasiCrys}.
 }\label{LevelScheme}
\end{figure}

Solitons are macroscopic quantum objects maintaining their integrity under self-trapped nonlinearities, with the 3D matter version being proposed by Rydberg dressing \cite{MauHen11}.   
Soliton molecules are bound states of solitons \cite{Str95, Gre02} that are macroscopic version of matter molecules with analogy in different properties such as vibration, synthesis, and dissociation \cite{Kru17}, providing a simulating lab for quantum chemistry. In telecom application, optical soliton molecules improve the information transfer rate beyond the fundamental threshold \cite{Str05,Roh12}.  Hence making stable soliton molecule has raised significant interest in the last decade \cite{Wan19,Gre12,Mal91,Akh97,Tan01,Gre02,Str05,Kru17,Her17,Bur99}. 
Interaction of solitons is effectively the interaction of their self-generated potential wells created under the nonlinear interaction. 
The interference type forces between solitons are extremely sensitive to their relative phases. This would hinder the formation of stable solitonic molecules, see Supp for more details.   
The RnD scheme is applied here for designing the phase independent self-generated potential featuring   an inner soft-core attraction supporting the 3D self-trapped soliton, an outer repulsive shell (barrier) preventing soliton fusion and a second attractive layer (well) used for completing the bound state resulting to giant stable soliton molecules. In this scheme,  the distance and size of individual solitons in the molecule can be controlled dynamically with laser adjustment.

Rydberg dressed atoms have been proposed for making droplet crystals \cite{Hen12,Sey19} where individual atoms in the lattice are  replaced by droplets containing many atoms. In the new arrangement, each atom only contributes in a small fraction of the droplet's energy. Therefore,  quantum tunneling of an atom between lattice sites is associated with a small energy cost. Tuning the interaction strength, droplet crystals  could reveal simultaneous superfluidity and crystalline order, forming an exotic quantum matter called super-solid  \cite{Hen12,Cin10,Cin14,Mac14,Osy11}.
 The current article shows that adding Gaussian peaks to the conventional plateau structure with  RnD scheme, diverges the exclusive triangular structure in droplet crystals to variety of quasi-periodic orders containing more than one tile for tessellation.
 These tiles in quasi-crystals  have n-fold rotational symmetry containing e.g.  n=5,7,8,9-fold symmetry that are forbidden in classical crystalline orders \cite{Lev84}.  Hence Quasi-crystals contain spatial patterns that are neither periodic as classical crystals nor totally disordered. 
Like its classical counterpart \cite{Gri03,Tan14,Koh86,Fuj91,Fuj91,Pie93,Wes03}, it is expected that quasi-ordering of droplet crystals to enrich the characteristics of this quantum matter.

\begin{figure}
 \scalebox{0.57}{\includegraphics*[viewport=30 110 500 550]{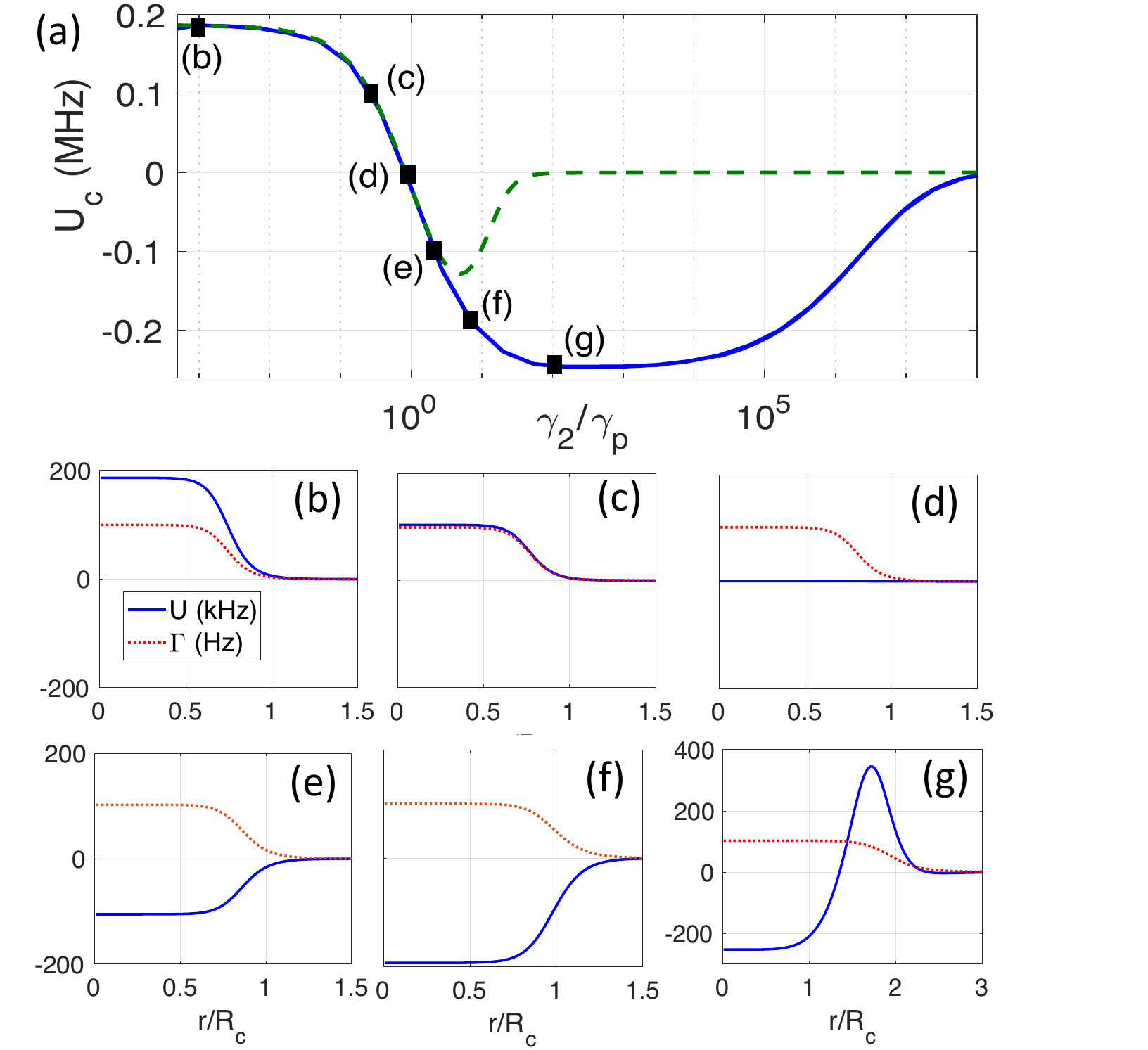}} 
\caption{ Switching between attractive and repulsive  interaction  with the noise knob. Fig (a) represents the evolution of soft-core interaction strength $U_c$ for the case with (blue solid line) and without (green dashed line) laser locking as a function of laser noise $\gamma_2$.  Here $\Omega_1$ is tuned to fix the loss rate per atom to $\Gamma$=100Hz. The noise rate of both lasers  are assumed to be equal ($\gamma_{1}=\gamma_{2}$). By increasing the laser de-phasing, soft-core interaction changes sign at $\gamma_{2}=\gamma_p$. 
(b-g) Spatial shape of RnD interaction is plotted in kHz (blue lines) for the marked laser de-phasing rates in (a). Dotted red lines show the  loss profile $\Gamma$ in Hz. Laser parameters are $\Delta/2\pi=10$MHz,  $\Omega_2/2\Delta=1$, targeting n=100 of Sr atoms. }\label{SignChange}
\end{figure}

An ultra-cold gas of N three-level strontium atoms that undergo in-resonance two photon excitation to the highly excited Rydberg state $|e\rangle$ is considered, see Fig 1a for the level scheme. 
The associated Hamiltonian for each atom can be written as
\begin{equation}
H_i=\frac{\Omega_1}{2} (\hat{\sigma}^i_{gp}+\hat{\sigma}^i_{pg})+\frac{\Omega_2}{2}  (\hat{\sigma}^i_{ep}+\hat{\sigma}^i_{pe})-\Delta \hat{\sigma}_{pp}
\end{equation}
where $\sigma_{\alpha,\beta}=|\alpha \rangle\langle \beta|$. 
The steady state of this system is a dark state $| d \rangle \propto \Omega_2 | g \rangle - \Omega_1 | e \rangle$ with zero light shift. In the limit of $\frac{\Omega_1}{\Omega_2}\ll1$ ground state atoms will be partially dressed with Rydberg states with population of $P_e=(\frac{\Omega_1}{\Omega_2})^2$.  
The effective two-body interaction $V_{ij}=C_6/r_{ij}^6 \sigma_{ee}^i \sigma_{ee}^j$ is a van-der Waals interaction between Rydberg atoms that is a function of interatomic distance $r_{ij}$. The interaction of $n\, ^3S_1$ Rydberg atoms is repulsive with van-der Waals coefficient being in the range of 120kHz.$\mu$m$^6<C_6< $73THz.$\mu m^6$  for principal numbers $24<n<100$. This strong interaction could exceed atom-light coupling over  several micrometers of interatomic separations.
Furthermore,  different decoherence terms  including spontaneous emission from Rydberg $\sqrt{\gamma_r }|p\rangle \langle e|$ and intermediate level $\sqrt{\gamma_p}|g\rangle \langle p|$ are encountered, where  $\gamma_{p}/2\pi=7.6$kHz in strontium and $\gamma_{r}$ can be found in \cite{Kun93}.
Effects of lasers' line-width are considered with de-phasing rates  $\gamma_{1}$ and $\gamma_{2}$  corresponding to 689nm ($\Omega_1$) and 319nm ($\Omega_2$) lasers respectively.  Effects of laser locking included  by $\gamma_{\text{Lock}}$,  ranging  between $|\gamma_{1}-\gamma_{2}|$ and $|\gamma_{1}+\gamma_{2}|$ for the cases of locking out of phase  up to the worst case of in-phase fluctuations. 
Finding the steady state of two interacting atoms $\rho_{ij}$ with the method explained in the supplementary,  the total interaction would be given by the sum of the binary interactions  
\begin{equation}
\bar{U}(r_{ij})=\text{Tr}[\hat{\rho}_{ij} (\hat{H}_i+\hat{H}_j+\hat{V}_{ij})].
\end{equation}
In the following figures, the background interaction independent light-shift $\bar{U}(r=\infty)$ that only generates a constant phase, is subtracted to present the effective interaction $U(r)=\bar{U}(r)-\bar{U}(\infty)$. 
The conventional Rydberg dressing interaction features plateau shape with the soft-core radius $R_c$, see Fig. 2. The soft-core radius is the distance at which the  interaction induced laser detuning equals the effective laser frequency bandwidth
$\mathrm{P_e}^2 V^2_{(\mathrm{R_{c}})}=(\frac{\Omega_1\Omega_2}{2\Delta})^2+\gamma_1^2+\gamma_2^2$.    The strong interaction within the soft-core makes the atoms out of resonance with the laser. Hence the atoms experience a collective light-shift generating a Kerr-type interaction \cite{Kha16, KhazDis16}. 

The  loss rate per atom is calculated as $\Gamma=\text{Tr}[\hat{\rho}_{i} (\gamma_p \hat{\sigma}_{pp}+\gamma_r \hat{\sigma}_{ee})]$. Rydberg interaction disturbs individual atom's dark state, populating short-lived intermediate state $|p\rangle$ within the soft-core radius, and hence increases the loss rate per atom within the soft-core, see Fig. 1b. Strong, un-locked laser de-phasing can also disturb dark state even for non-interacting atoms. However, locking the lasers could protect the dark state even for de-phasing rates close to EIT bandwidth, see  Fig.  \ref{LaserLock}c in Supp.
While de-phasing is usually treated as a destructive effect, here  laser noise is used as a controlling parameter to manipulate the strength and design the spatial shape of interaction. Fig. \ref{SignChange} shows that for a constant loss rate per atom $\Gamma=100$Hz, increasing the  laser noise $\gamma_2$ could result to a sign change that happens at $\gamma_2=\gamma_p$, see Fig.  \ref{SignChange}a. More detailed discussions about the sign-change can be found in the supplementary. 
Fig. \ref{LaserLock}a shows that with locking the lasers, interaction would survive at de-phasing rates much larger than EIT bandwidth $\delta_{\text{EIT}}=\frac{\Omega_2^2}{4|\Delta|}$ \cite{Sev11,Gor11,Gau16}. This extra window is valuable for designing the interaction shape, see below.

\begin{figure}[!h]
\centering
    \scalebox{0.42}{\includegraphics*[viewport=0 55 600 550]{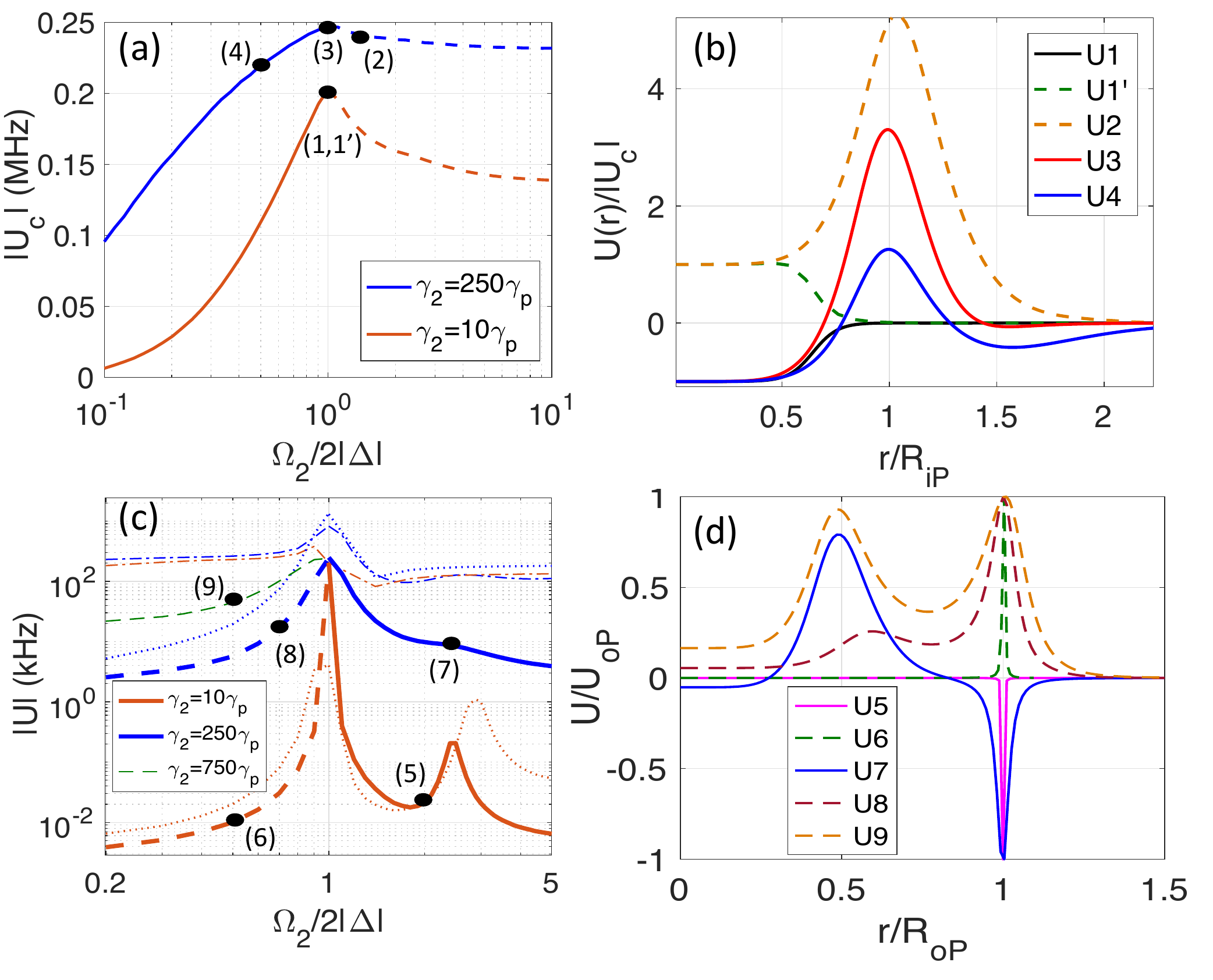}} 
\caption{Controlling interaction shape with laser noise.
(a,c) The scaling of interactions' magnitude of the soft-core $|U_c|$ are presented by thick solid (dashed) lines associated with positive (negative) detuning. The maximum loss rate over the profile is limited to $\Gamma=100$Hz. 
(c) The interaction magnitude of the (inner-) outer-peak ($|U_{iP}|$) $|U_{oP}|$ is shown by (dotted) doted-dashed lines. 
The interaction profiles for the marked points are presented in (b,d).
 Here $\Delta/2\pi=10$MHz, n=100,  $\gamma_1=\gamma_2$ and lasers are locked. To fix the maximum loss,  the lower laser is controlled within the range of $0.1<\Omega_1/2\pi<2$MHz. }\label{Potential}
\end{figure}

Enriching the interaction profile with extra peaks, extends the applications with examples provided below. 
 Applying adequate laser de-phasing   generates a first peak at inner-peak distance given by 
$V^2(\mathrm{R_{iP}})=\Delta^2+\frac{\gamma_2+\gamma_p+\gamma_r}{4\gamma_r}\Omega_2^2+\frac{(\gamma_2+\gamma_p+\gamma_r)^2}{4}$, see Fig. \ref{Potential}b.
The effective line-broadening, results to blockade leakage. This would result to the increase of two-atom Rydberg excitation $\rho_{ee,ee}$ and lead to direct Rydberg interaction  $\frac{C_6}{r^6} \rho_{ee,ee}$, see Fig. \ref{DensityMatrix}b. Note that blockade leakage only increases $\rho_{ee,ee}$, leaving Rydberg and the intermediate state populations $\rho_{ee}$, $\rho_{pp}$ unaltered at the regime of interest. Therefore, the loss does not increase over this peak.
By reducing $\Omega_2/2\Delta$ or increasing $\gamma_2$, the soft-core could extend beyond the inner-peak, making valuable profile like U4 in Fig. \ref{Potential}b. In this case, the peak effectively splits the soft-core making an  inner soft-core radius of $\mathrm{R_{iC}=R_c}$ and the outer radius given by $V(\mathrm{R_{oC}})=2\delta_{\text{EIT}}$.

The second peak is formed  by the sudden increase in the intermediate level's population $\rho_{pp}$ and hence the loss rate increases at the outer-peak position, see Fig. \ref{DensityMatrix}c in the supplementary.
This peak happens in two regimes of parameters ($\Omega_2>2|\Delta|$ with $\Delta>0$) and ($\Omega_2<2|\Delta|$ with $\Delta<0$).
The position of the outer-peak would be at
\begin{equation}
V(\mathrm{R_{oP}})=\frac{2\Delta \Omega_2^2}{(\Omega_2^2+\gamma_p^2-4\Delta^2)},
\end{equation}
see Fig. \ref{Potential}d. 

The main features of RnD interaction profile are discussed in  Fig. 3,  for the case that the maximum loss over the profile is fixed to $\Gamma=100$Hz. 
The soft-cores and outer-peak are formed by the interaction induced collective light-shift. Therefore, the interaction sign in these elements are defined by the sign of $\Delta$ and the value of $\gamma_2$. Being interested in the parameter regime $\gamma_2\gg \gamma_p$ in Fig. 3, the positive (negative) interaction is generated  by $\Delta<0$ ( $\Delta>0$) shown by dashed (solid) lines in figure 3. The inner-peak is directly generated by the van-der Waals interaction and hence its sign only follows the sign of $C_6$.
In Fig. 3, the scaling of interactions' magnitude at the soft-core $|U_c|$ are presented by thick solid/dashed lines for the case of positive/negative detuning. One can see that there is an optimum interaction to loss ratio $U_c/\Gamma=2000$ at $\Omega_2/2\Delta=1$   providing an opportunity to observe quantum mechanical effects in ultra-cold atoms, see below.
 Interaction strength of inner- and outer-peak $|\mathrm{U_{iP}}|$ and $|\mathrm{U_{oP}}|$ are presented by dotted and dotted-dashed lines respectively. As  can be seen, the magnitude of $|\mathrm{U_c}|$ and $|\mathrm{U_{iP}}|$ strongly depend on $\gamma_2$ and $\Omega_2$  while the outer-peak  mainly depends on the detuning at the regime of interest. This feature could be used to isolate  the outer-peak,  forming a shell type interaction profile in 3D. This type of interaction profile is ideal for controlling interacting sites in optical lattices with global lasers. 

\begin{figure}[!h]
\centering
    \scalebox{0.52}{\includegraphics*[viewport=0 00 500 530]{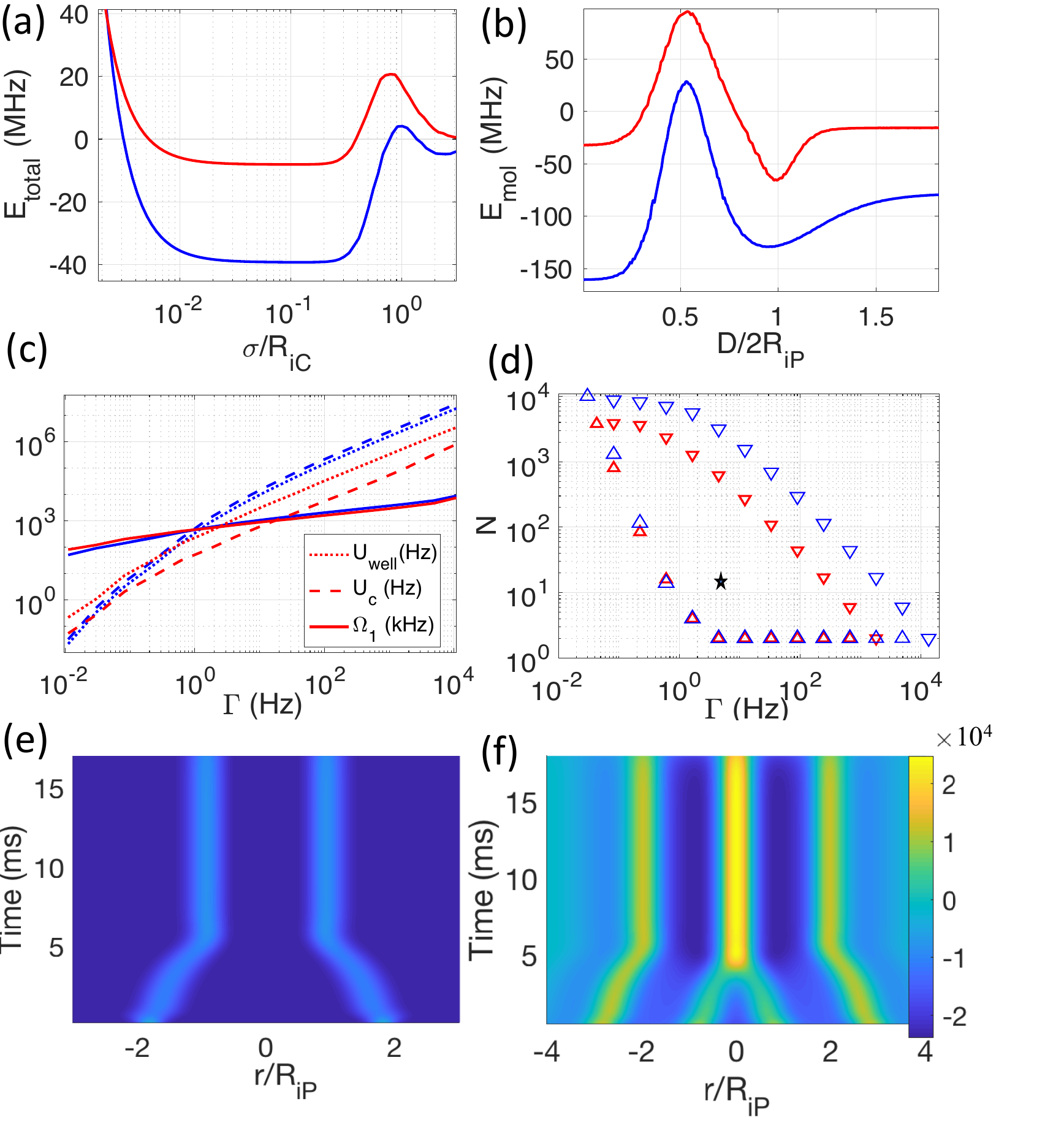}} 
\caption{Stability of 3D soliton molecules formed by the red/blue potentials of Fig. 1b, distinguished by the same color coding.  Total energy of a single soliton  (a bi-soliton molecule) with 20 (40) atoms is plotted as a function of HWHM of soliton's Gaussian profile $\sigma$  (inter-solitonic distance $D$) in Fig. a (b). The same interaction profiles of Fig. 1b, would scale in magnitude by manipulating $\Omega_1$.
 The scaling of soft-core (well) interaction $U_c$ ($U_{well}$) as a function of the maximum loss rate over the profile $\Gamma$ and the corresponding $\Omega_1$ are plotted in (c) with $2.5<R_{\text{iC}}<5\mu$m. (d) The minimum  (maximum) atom numbers required for stable molecule formation are plotted by upward (downward) triangles as a function of $\Gamma$, see the text for more details. 
Molecule synthesis with parameters shown by star sign in (d), are  presented as the time evolution of (e) $|\psi|^2$ and (f) the mean-field interaction $W$. 
Laser parameters in red (blue) curves of Fig. 1b are $\gamma_1=\gamma_2=100(1000)\gamma_p$, $ \Omega_1/2\pi=450(330)$kHz,  $\Omega_2/2\Delta=2.5(0.3) $, $\Delta/2\pi=2(10)$MHz, n=100, and the lasers are locked out of phase. Solitons' HWHM in (b-d) are $\sigma=0.2R_{\text{iC}}$.
 }\label{MolFusibility}
\end{figure}

 The designed red/blue RnD interactions shown in Fig. 1b are ideal for making giant stable soliton molecules. The inner soft-core results to 3D soliton formation.
 The positive inner-peak (barrier) preserves the integrity of molecule's elements over the synthesis process or in case of strong vibration.  To avoid the conventional phase dependent interaction, an outer-shell of potential well is developed in the interaction profile outside the barrier for inter-solitonic attraction.
Stability of molecule formed by the red/blue potentials of Fig. 1b are discussed in Fig. \ref{MolFusibility}.  Figure \ref{MolFusibility}a,b represent  the total energy of a single and a molecule solitons  as a function of individual soliton's Gaussian HWHM and inter-solitonic distance respectively. Dressing energy $E_{\text{dress}}=\sum\limits_{i<j}U(r_{ij})$ is numerically calculated using Marsaglia polar method for Gaussian random sampling averaging over 100 trials. 
Adjusting  loss and interaction strength by manipulating $\Omega_1$ do not alter interaction's spatial profile. 
The scaling of soft-core (well) interaction $U_{\text{c}}$ ($U_{\text{well}}$) as a function of the maximum loss rate over the profile $\Gamma$ and the corresponding value of $\Omega_1$ are plotted in Fig. \ref{MolFusibility}c.
The range of allowed atom numbers in the molecule shown in Fig. \ref{MolFusibility}d, are determined by stability and decay rates.  
The minimum atom number $N_{\text{min}}$ (upward triangle) is defined by the condition that the energy of attractive forces ($E_{\text{dress}}\propto N^2$) must overcome the one of repulsive forces ($E_{k}\propto N$). 
The maximum atom number $N_{\text{max}}$ (downward triangle) is defined by the condition of not losing any atom with 37\% Poisson probability  over 2 molecule oscillation cycles period $t_2$ (i. e. $N_{\text{max}}\Gamma t_2=1$). Molecule oscillation frequency is defined by $\omega_{\text{mol}}=\sqrt{\frac{\hbar U_{\text{mol}}}{N m \sigma_{\text{osc}}^2}}$, where $U_{\text{mol}}$ and  $\sigma_{\text{osc}}$ are the depth and the HWHM of the molecule potential well and $N\times m$ is the molecule mass. The main advantage of blue potential that results to larger molecule mass, comes from the controlled loss rate over the attractive well.

The evolution of molecule synthesis generated under blue potential is simulated in Fig. \ref{MolFusibility}e. 
 In the zero temperature limit the dynamics of the Rydberg dressed BEC is governed by Gross-Pitaevskii equation (GPE), 
\begin{equation}
\text{i} \hbar \partial_t \psi(x)=[-\frac{ \hbar^2 \nabla^2}{2m} + g |\psi|^2 + W(x)] \psi(x)
\end{equation}
where $W(x)= \int U(x-x') |\psi(x')|^2 \text{d}x'$ is the mean field interaction caused by non-local  long range RnD interaction  $U$
 and $g=\frac{4\pi \hbar^2 a}{m} $ is the local interaction caused by s-wave scattering of atoms with scattering length of $a=96a_0=5088$pm \cite{Ste14}.
 There are  $|\langle \psi | \psi \rangle |^2=N$ strontium atoms with mass $m$ in the condensate.
The initial state of two solitons distinguished by left ($l$) and right ($r$) indices is given by $\psi_0(x)=A_r\sech(\sigma_r(x-x_0)) \text{e}^{\text{i} v_r x}+A_l\sech(\sigma_l(x+x_0)) \text{e}^{\text{i} v_l x}$ that is normalized to particle numbers  $\int |\psi_0|^2 \text{d}x=N$.
To avoid phase dependent interaction, solitons are confined to the inner soft-core $R_{\text{iC}}$ leaving no population in the outer well.

\begin{figure}[!h]
\centering 
      \subfloat{%
    \includegraphics[width=.24\textwidth]{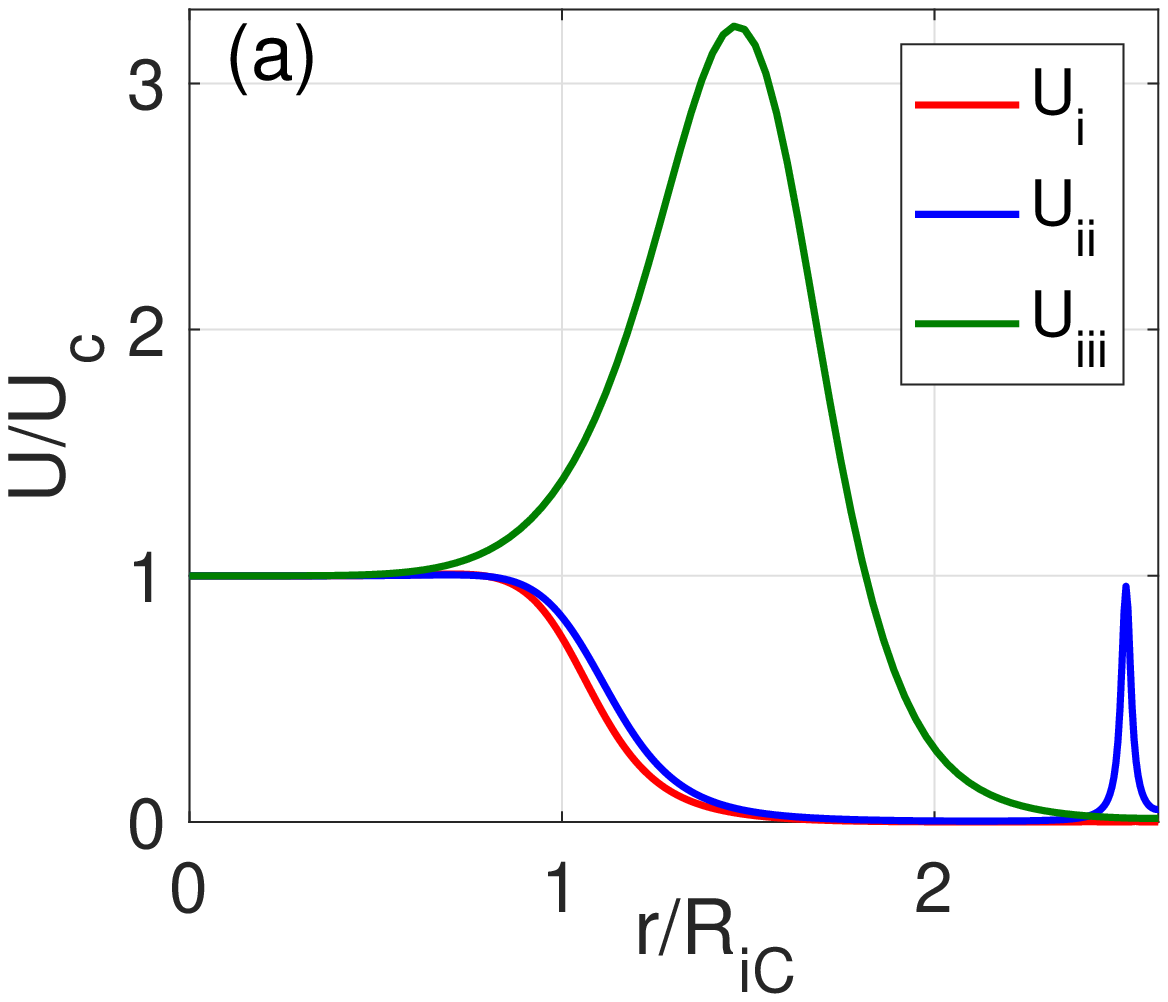}}\hfill       
    \subfloat{%
    \includegraphics[width=.24\textwidth]{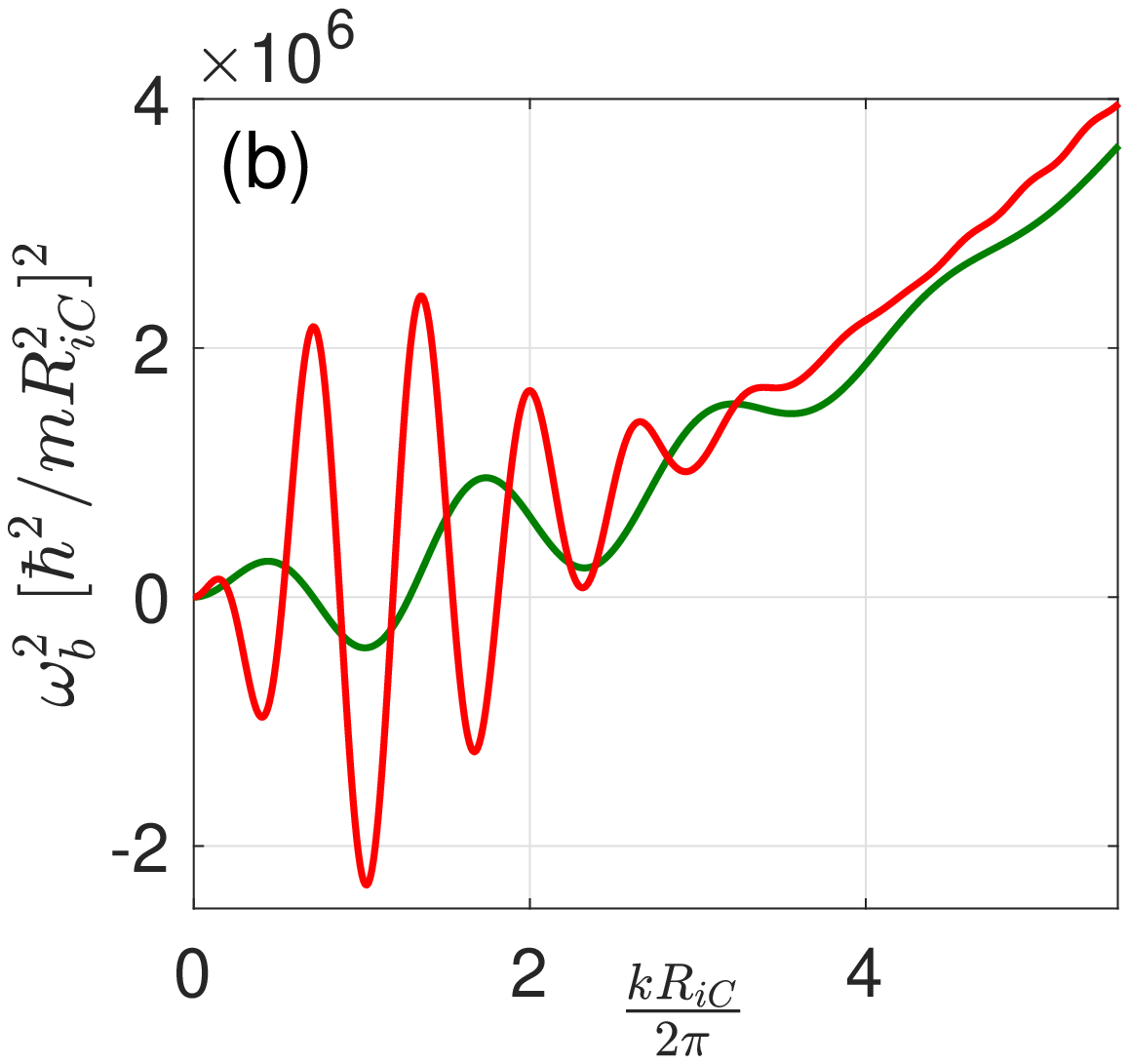}}\hfill 
          \subfloat{%
    \includegraphics[width=.24\textwidth]{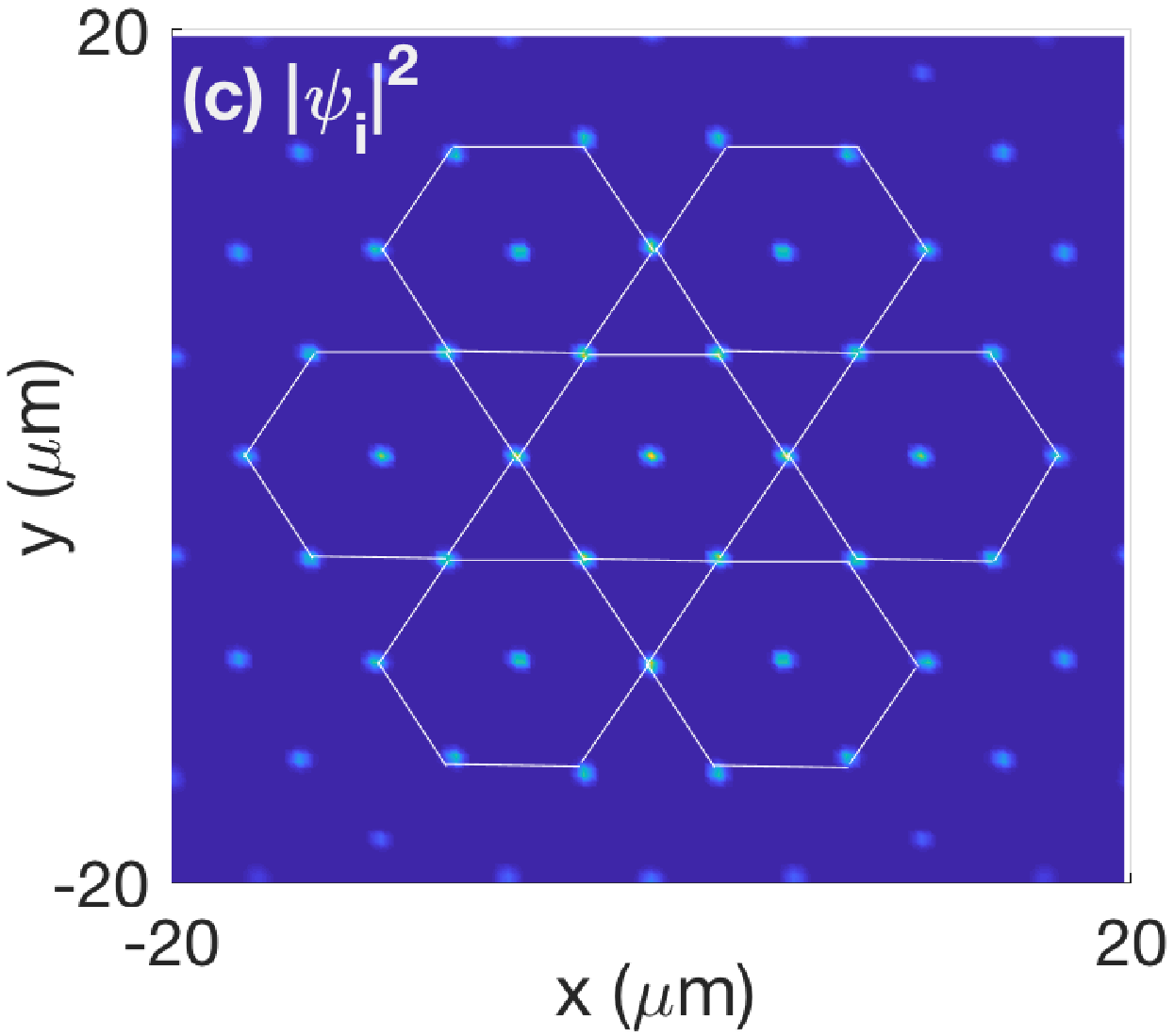}}\hfill 
          \subfloat{%
    \includegraphics[width=.24\textwidth]{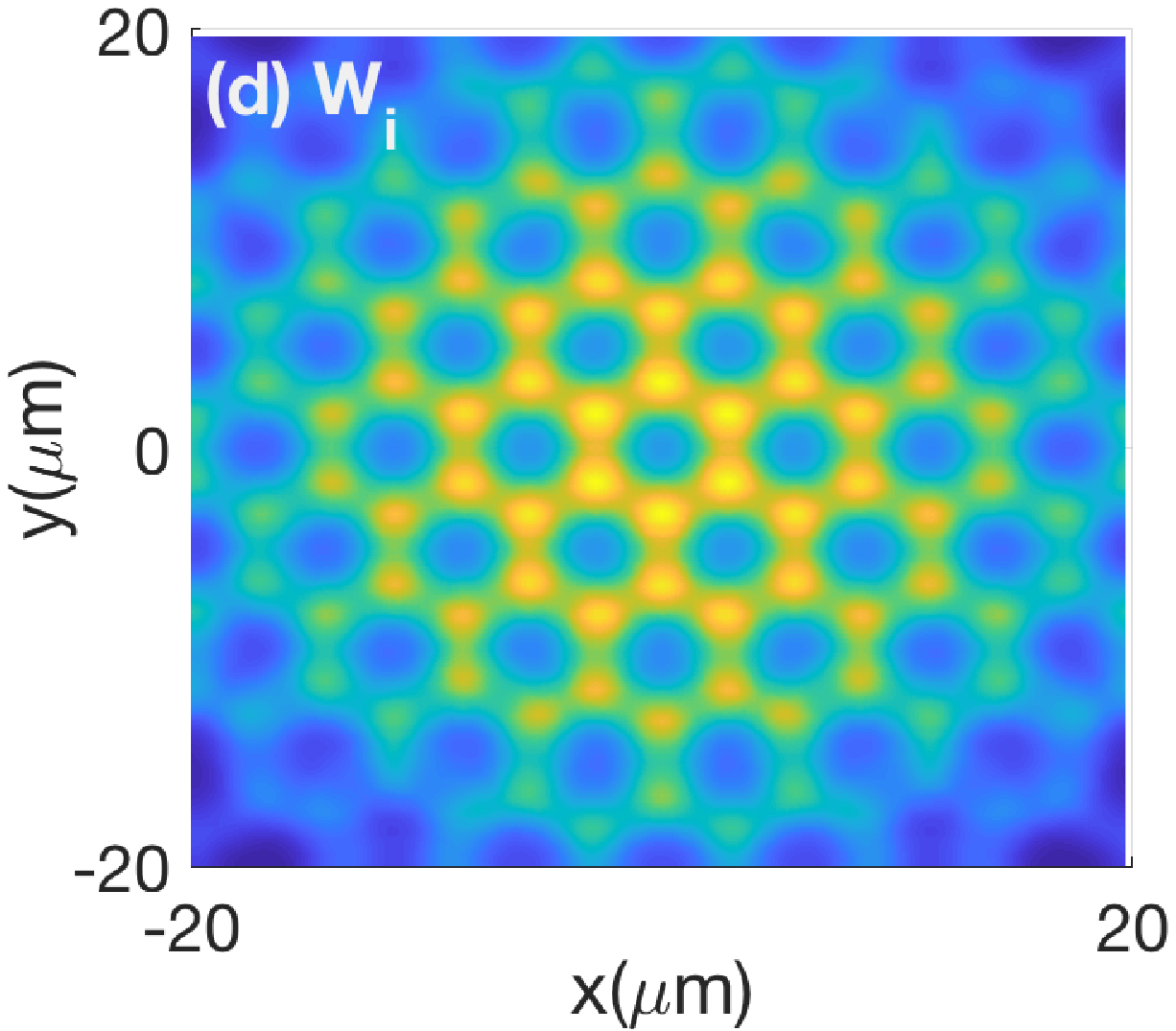}}\hfill 
          \subfloat{%
    \includegraphics[width=.24\textwidth]{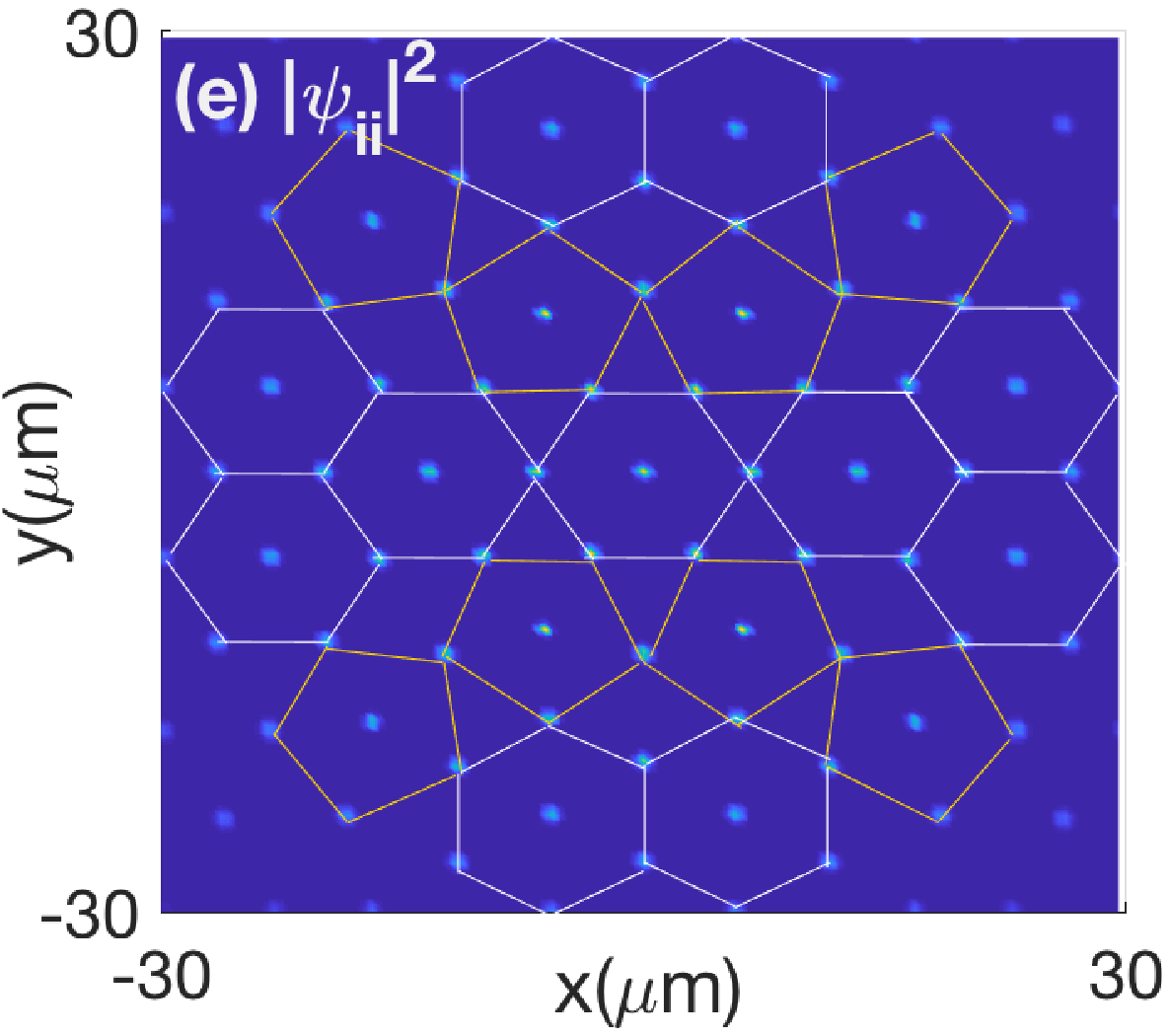}}\hfill 
          \subfloat{%
    \includegraphics[width=.24\textwidth]{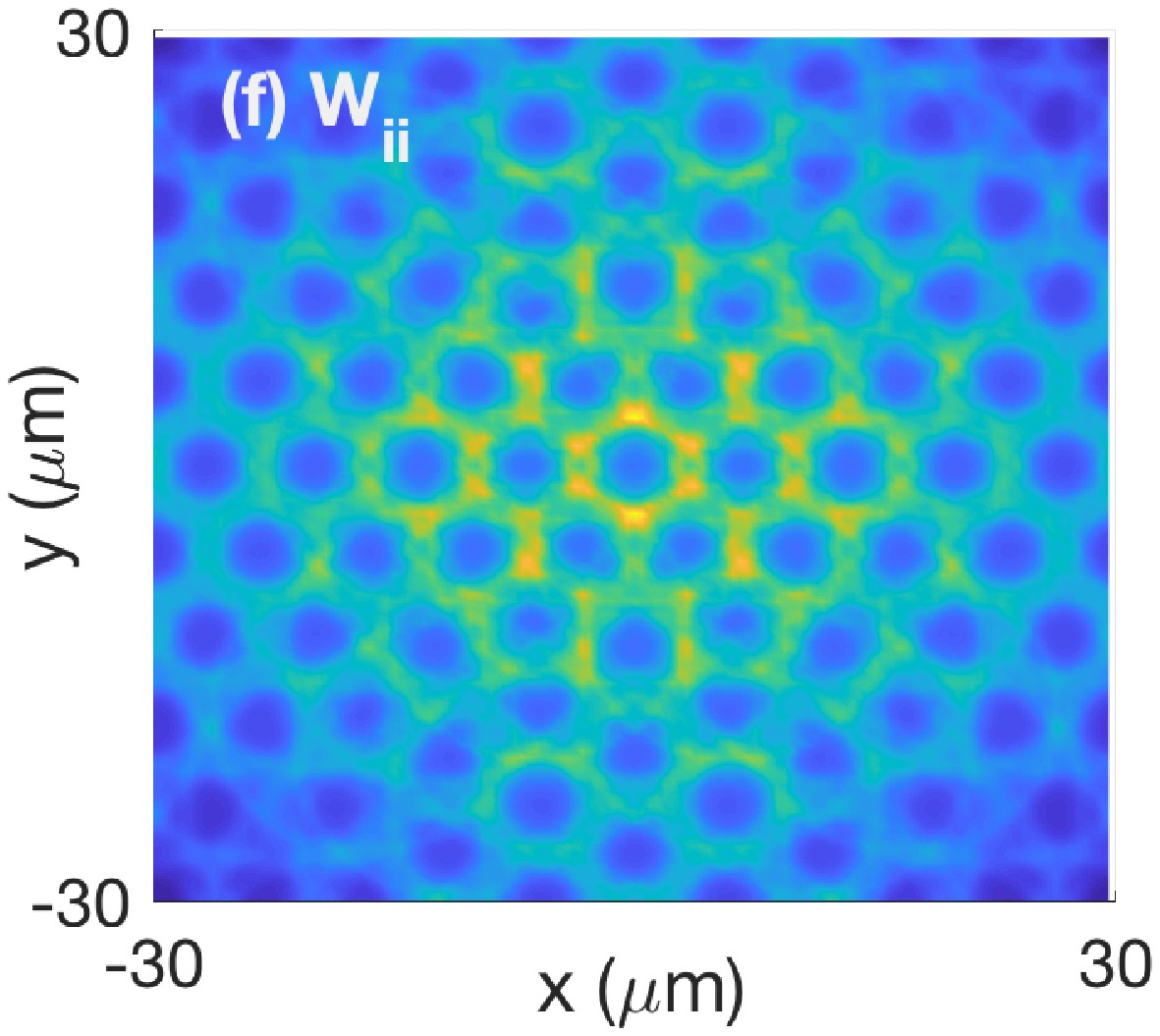}}\hfill 
            \subfloat{%
    \includegraphics[width=.24\textwidth]{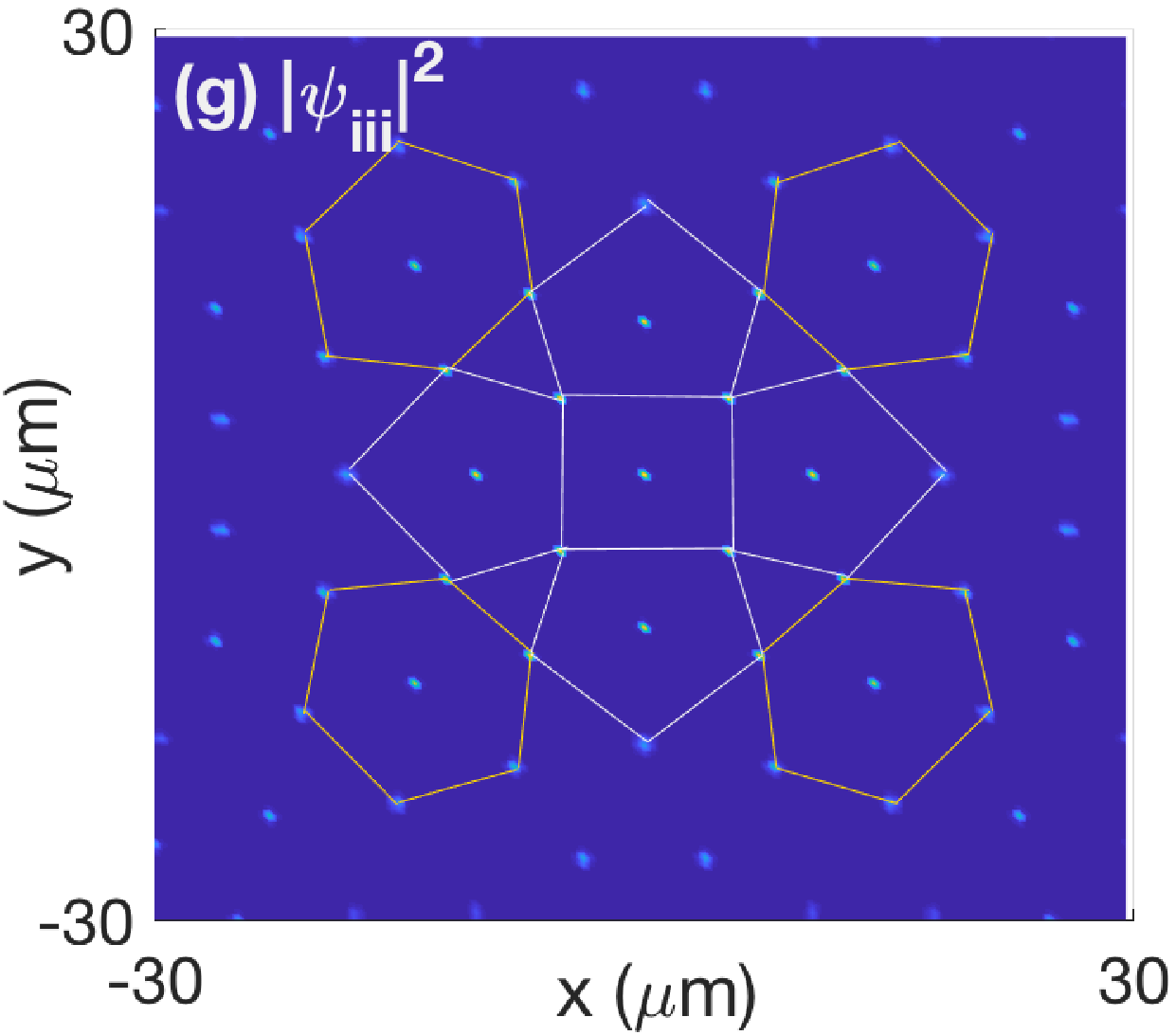}}\hfill 
            \subfloat{%
    \includegraphics[width=.24\textwidth]{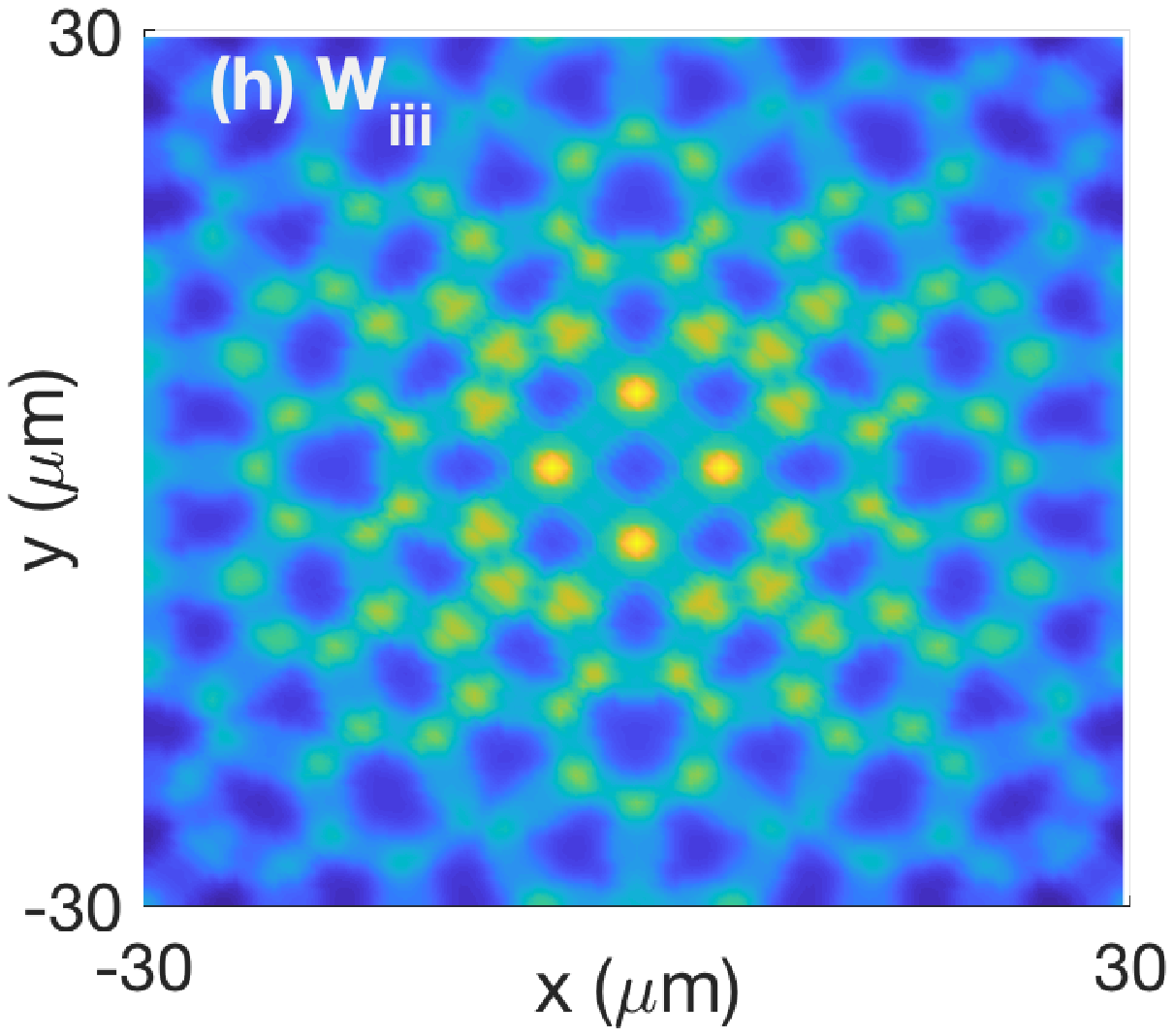}}\hfill 
\caption{Droplet quasi-crystals. (a,b)  Adding a Gaussian to the conventional plateau in RnD interaction profiles, results to more than one roton unstable points in Bogoliubov spectrum (b). Hence there will be more than one lattice constants leading to quasi-ordered crystals (e-h). (c-h) The ground state of 2D GPE under potentials $U_i$-$U_{iii}$ are plotted as $|\psi|^2$ (left column) and the mean field interaction $W$ (right column). Dynamic parameters of $U_i$-$U_{iii}$ are $\Delta/2\pi=10$MHz,  $\Omega_2/2\Delta=1$, $\Omega_1/2\pi=[145, 145, 260]$kHz,  n=100, $\rho$=10$^{13}$cm$^{-3}$, $\gamma_1=\gamma_2=[10,10,100]\gamma_p$, and lasers are not locked exclusively in $U_{ii}$. }\label{QuasiCrys}
\end{figure}

 In a different application, adding a Gaussian peak to the conventional plateau profile  in RnD interaction could  be used for enriching the  patterns in droplet crystals that was limited to triangular pattern.   
 This could be seen by looking at the  time evolution of the  quasi-particle excitations $\delta \psi_{\bf k}({\bf r})=u_k \text{e}^{\text{i}{\bf k.r}-\text{i}E_b(k)t}+\bar{v}_k \text{e}^{-\text{i}{\bf k.r}+\text{i}E_b(k)t}$ that is determined by Bogoliubov spectrum 
$\omega_b^2(k)=\frac{\hbar^2 k^2}{2m} (\frac{\hbar^2 k^2}{2m} + 2\rho g+ 2\rho \tilde{U}(k))$
where $\tilde{U}(k)$ is the Fourier transform of dressing interaction and $\rho$ is the density of atoms. The strong soft-core interaction in Rydberg dressed atoms results to roton unstable point with imaginary spectrum  $E_b(k_{\text{roton}})=i\beta$. Therefore, the excitation around roton instability $k_{\text{roton}}$ would grow exponentially with the rate $\beta$. This would form a spatial periodic structure with lattice constant of $a=\frac{2\pi}{k_{\text{roton}}}$.  The soft-core in conventional Rydberg dressing forms a single roton unstable point and hence generates a triangular matrix \cite{Hen12}. In this work, adding a Gaussian to the soft-core in the RnD interaction results to multiple roton instabilities forming more than one lattice constants, see Fig. \ref{QuasiCrys}b . This would result to quasi-periodic structured droplet crystal formation.
To find the ground state, imaginary time evolution of Gross Pitaevskii equation (GPE) is numerically calculated in Fig. \ref{QuasiCrys}[e-h] for the potentials of Fig \ref{QuasiCrys}a. While the plots are presented in semi 2D with a disc trap arrangement $\omega_z>>\omega_r$, RnD interaction  could be used for making 3D quasi-crystals as well.

{\bf Outlook:} This article shows that Rydberg dressing of atoms with noisy lasers could change the sign of interaction as well as adding new features to interaction profile. 
The possibility of changing interaction sign opens new insights for the current interactions in AMO physics with exclusive attractive or repulsive nature e.g.  \cite{Zha18}.
Designing a hybrid potential profile, this article showed that RnD could be used for realizing  3D stable soliton-molecules of 10$^4$ atoms. The  level of control over the physical elements and exaggerated size and mass of these coherent molecules provide an ideal case for the study of  molecules'  chemistry in the transition from microscopic to the macroscopic regime.
Presented RnD potentials in Fig. \ref{Potential} could also  be used to test the quantum collision of heavy coherent systems at small speeds that is currently demanding to close the experimental hole in testing collapse models \cite{Tor18,Car19}.  
Considering the vast studies in the field of Rydberg photonics \cite{Ada19}, extending the current matter-molecule scheme to optical system is trivial with valuable applications in telecommunications.

{\it Acknowledgment:} The author thanks  T. Pohl and Y. Zhang for fruitful discussions and helpful comments.

\section{Supplemental material}

\subsection{In-resonance Rydberg dressing with noisy lasers}

Dynamics of the two-body density matrices $\hat{\rho}_{ij}=\text{Tr}_{\bar{i},\bar{j}}\hat{\rho}$ that is traced over all but $i$ and $j$ particles, is given by
\begin{eqnarray}\label{Heisenberg}
\partial_t \hat{\rho}_{ij}=&&-\text{i}[\hat{H}_i+\hat{H}_j,\hat{\rho}_{ij}] + \mathcal{L}_i(\hat{\rho}_{ij})+\mathcal{L}_j(\hat{\rho}_{ij}) \\ \nonumber
&&- \text{i}[\hat{W}_{ij},\hat{\rho}_{ij}] - \text{i} \sum \limits_{k \neq i,j}\text{Tr}_k [\hat{W}_{ik}+\hat{W}_{jk},\hat{\rho}_{ijk}].
\end{eqnarray}
The last term contains three-body density matrices. Since the effective interaction is limited in space $r_{ik},r_{jk} \lesssim R_c$, population of Rydberg atoms in the interaction region serves as a small number in the weak dressing regime. As a result, the interaction terms of the three-body density matrices would be  $\epsilon=P_eN_c$ times smaller than the  one in two-body density matrices and could be neglected ($N_c$ is the number of atoms within the soft-core radius $R_{\text{oC}}$).  

\begin{figure}
\centering
 \subfloat[]{%
    \includegraphics[width=.16\textwidth]{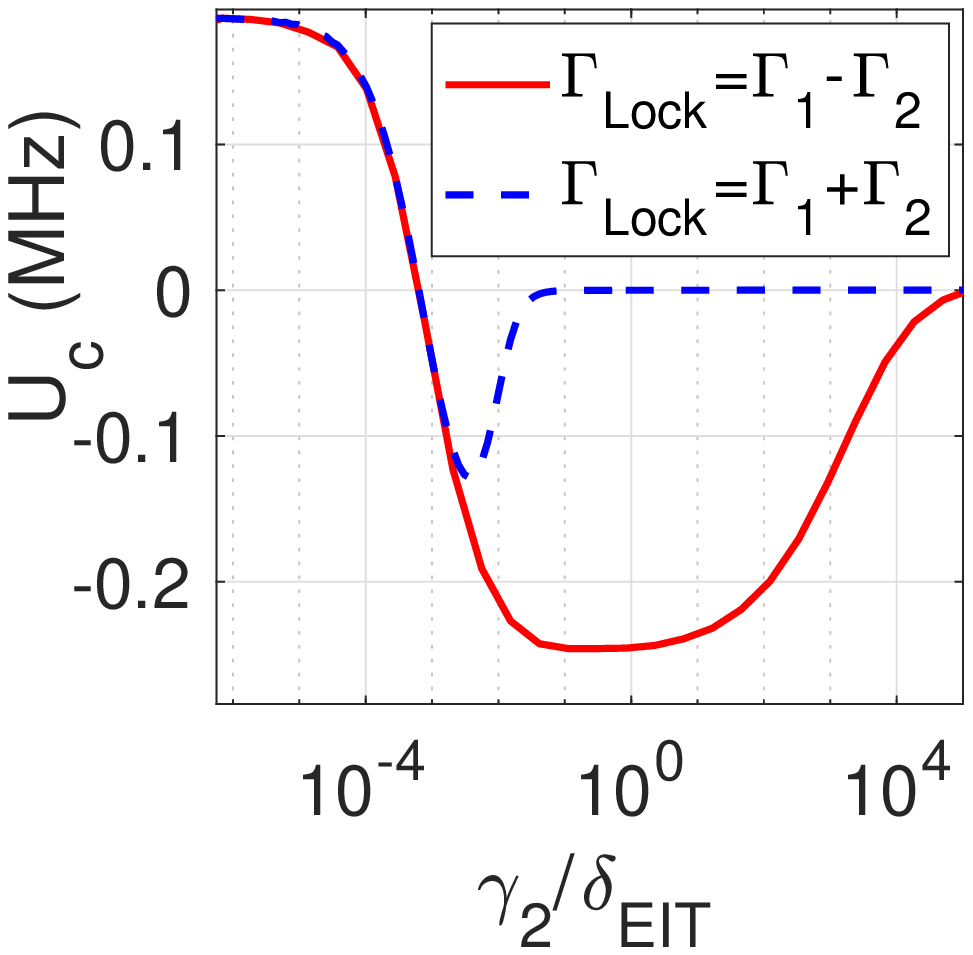}}\hfill 
      \subfloat[]{%
    \includegraphics[width=.16\textwidth]{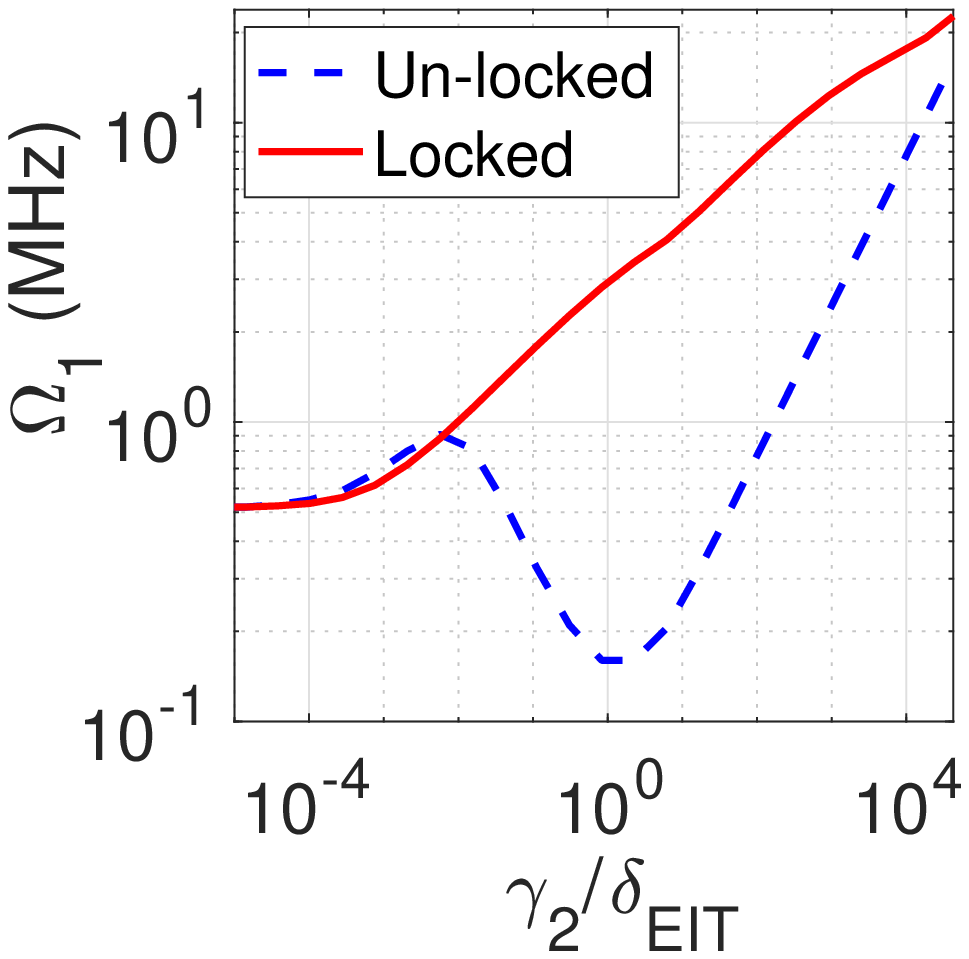}}\hfill 
       \subfloat[]{%
    \includegraphics[width=.16\textwidth]{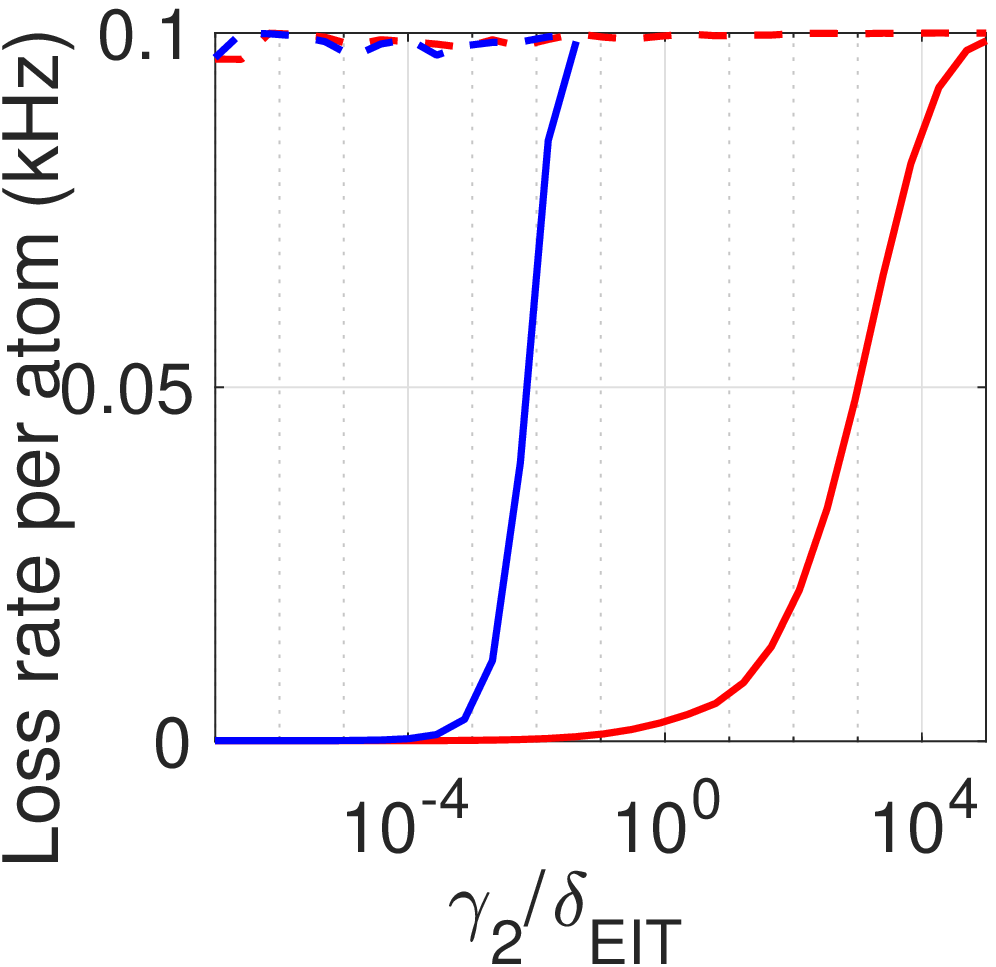}}\hfill  
\caption{ Effects of laser locking. Fig (a) represents the evolution of the soft-core interaction strength for the cases with and without laser locking as a function of laser noise.   The noise rate of both lasers  are assumed to be equal ($\gamma_{1}=\gamma_{2}$).  Laser locking provides a wider window over laser de-phasing rates to operate. (b) $\Omega_1$ is tuned to fix the loss per atom to 100Hz. (c) Interaction dependent and independent loss rates are plotted with dashed and solid lines respectively.  The cases of locked and unlocked lasers are distinguished by red and blue colors. Fig. (c) shows that laser locking prevents background loss (interaction independent loss) over wider range of laser de-phasing.  Applied parameters in this figure are $\Delta/2\pi=10$MHz, $\Omega_2/2\Delta=1$, n=100. }\label{LaserLock}
\end{figure}

Lindblad terms encounters  spontaneous emission from Rydberg $\sqrt{\gamma_e }|p\rangle \langle e|$ and intermediate level $\sqrt{\gamma_p}|g\rangle \langle p|$. Furthermore, laser fluctuations are included as  $\sqrt{\Gamma_{gg}}|g\rangle \langle g|$, $\sqrt{\Gamma_{pp}}|p\rangle \langle p|$, $\sqrt{\Gamma_{ee}}|e\rangle \langle e|$,
  where $\Gamma_{gg}=(\gamma_{\text{Lock}}+\gamma_{1}-\gamma_2)/2$, $\Gamma_{pp}=(-\gamma_{\text{Lock}}+\gamma_{1}+\gamma_2)/2$, $\Gamma_{ee}=(\gamma_{\text{Lock}}-\gamma_{1}+\gamma_2)/2$ with $\gamma_{1}$ and $\gamma_{2}$ being the line-widths of $\Omega_1$ and $\Omega_2$ lasers (corresponding to 689nm and 319nm lasers respectively). The laser locking effects are encountered in  $\gamma_{\text{Lock}}$ that has a value  between $|\gamma_{1}-\gamma_{2}|$ and $|\gamma_{1}+\gamma_{2}|$ for the cases of locking out of phase  up to the worst case of in-phase fluctuations. By locking the lasers, the dark state dressing and the desired interaction would be preserved over a wider range of laser noise even above EIT bandwidth, see Fig. \ref{LaserLock}.   
  
 \subsection{Interaction sign-change}
 
\begin{figure}
 \scalebox{0.33}{\includegraphics*[viewport=0 0 740 320]{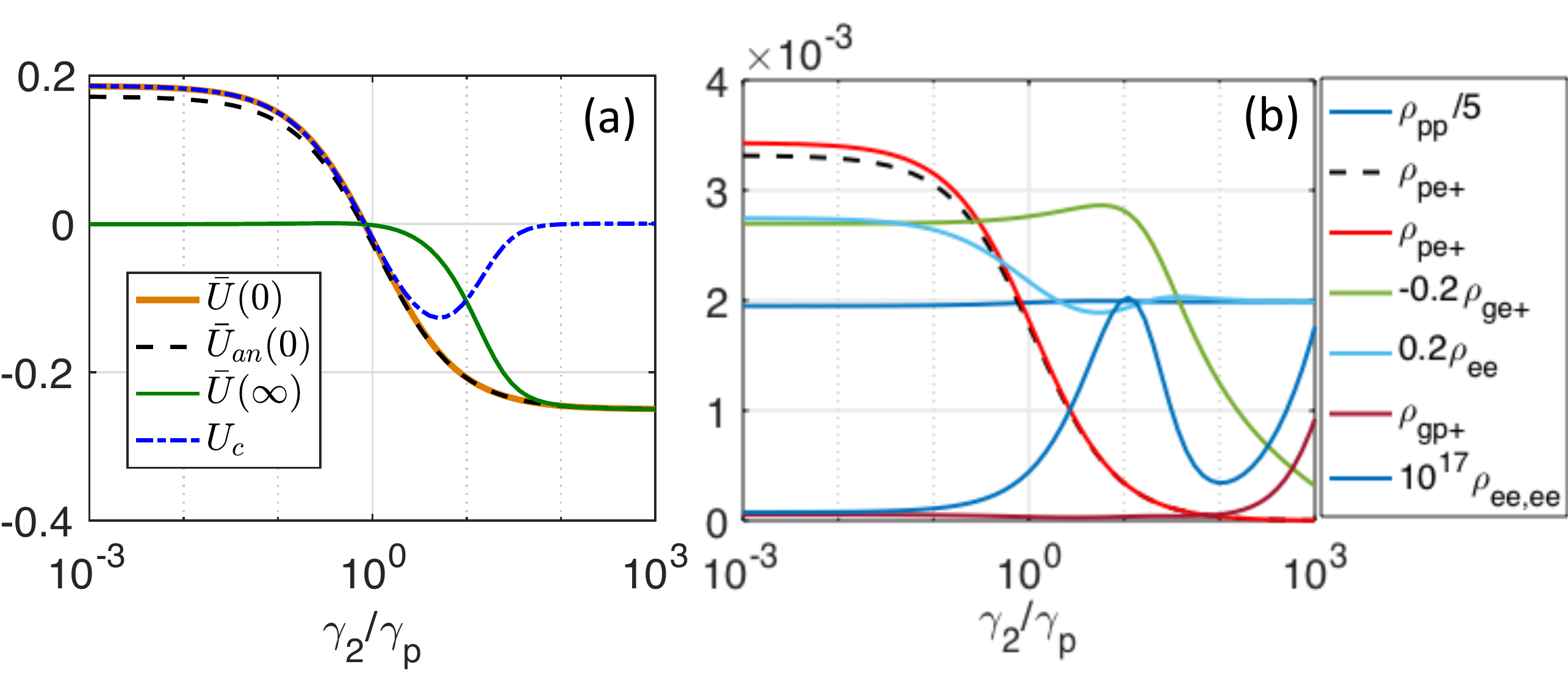}} 
\caption{ Interaction sign-change with laser noise knob. (a) To find the effective soft-core interaction $U_c$ (blue dotted-dashed line),  the light-shift of non-interacting atoms $\bar{U}(\infty)$ (green line) is subtracted from the light-shift of  interacting atoms $\bar{U}(0)$ (orange line).  The loss rate of interacting atoms is fixed to $\Gamma=100$Hz by adjusting $\Omega_1$. This would make $\rho_{pp}$ of interacting atoms to be constant.
(b) Interaction sign-change occurs due to the reduction of $\rho_{pe_{+}}$. Further increase of de-phasing $\gamma_2$ would destroy EIT, reducing $\rho_{ge}$ making effective interaction $U_c=0$. 
Black dashed lines in a(b),  presents $\bar{U}_{an}$ ($\rho_{pe_{+}}$) plotted based on Eq. 6(7).
Un-locked Lasers of $\Delta/2\pi=10$MHz, $\Omega_2/2\Delta=1$, $\gamma_1/2\pi=50$kHz are used to dress Sr atoms to $|5s100s ^3S_1\rangle $ Rydberg level.
}\label{SignChangeRho}
\end{figure}

\begin{figure}[!h]
\centering
 \subfloat[]{%
    \includegraphics[width=.24\textwidth]{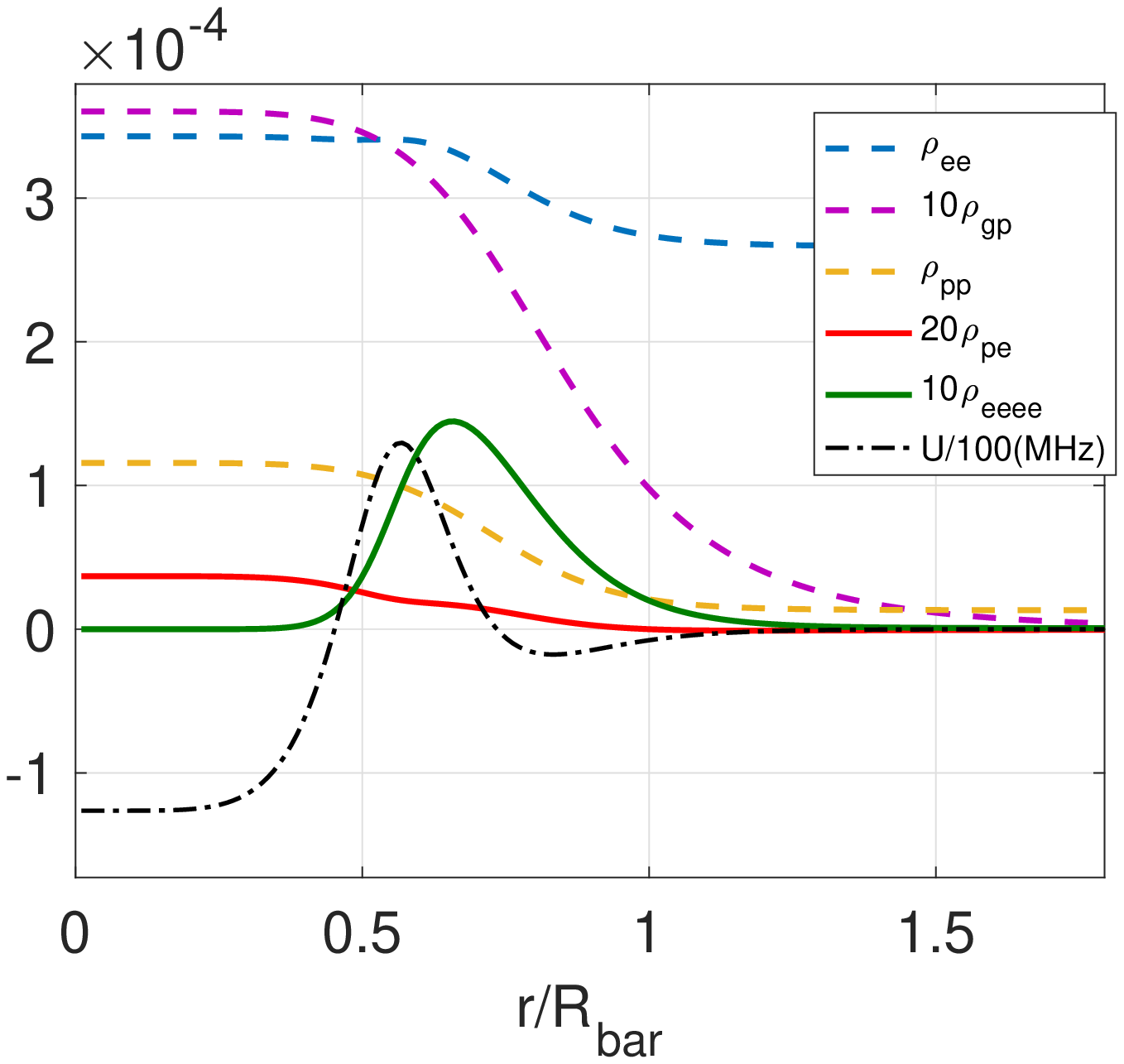}}\hfill 
 \subfloat[]{%
    \includegraphics[width=.24\textwidth]{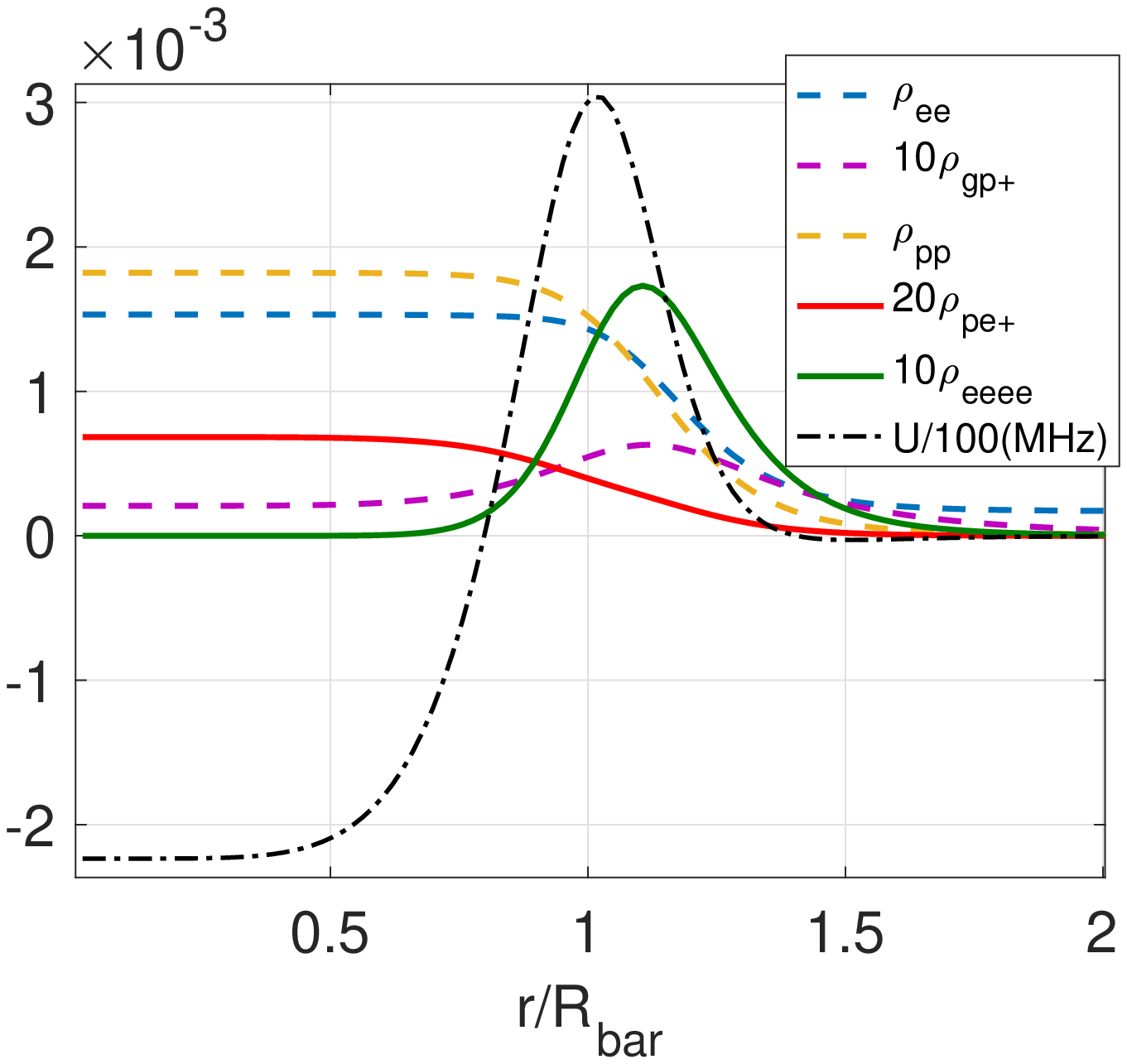}}\hfill 
     \subfloat[]{%
    \includegraphics[width=.24\textwidth]{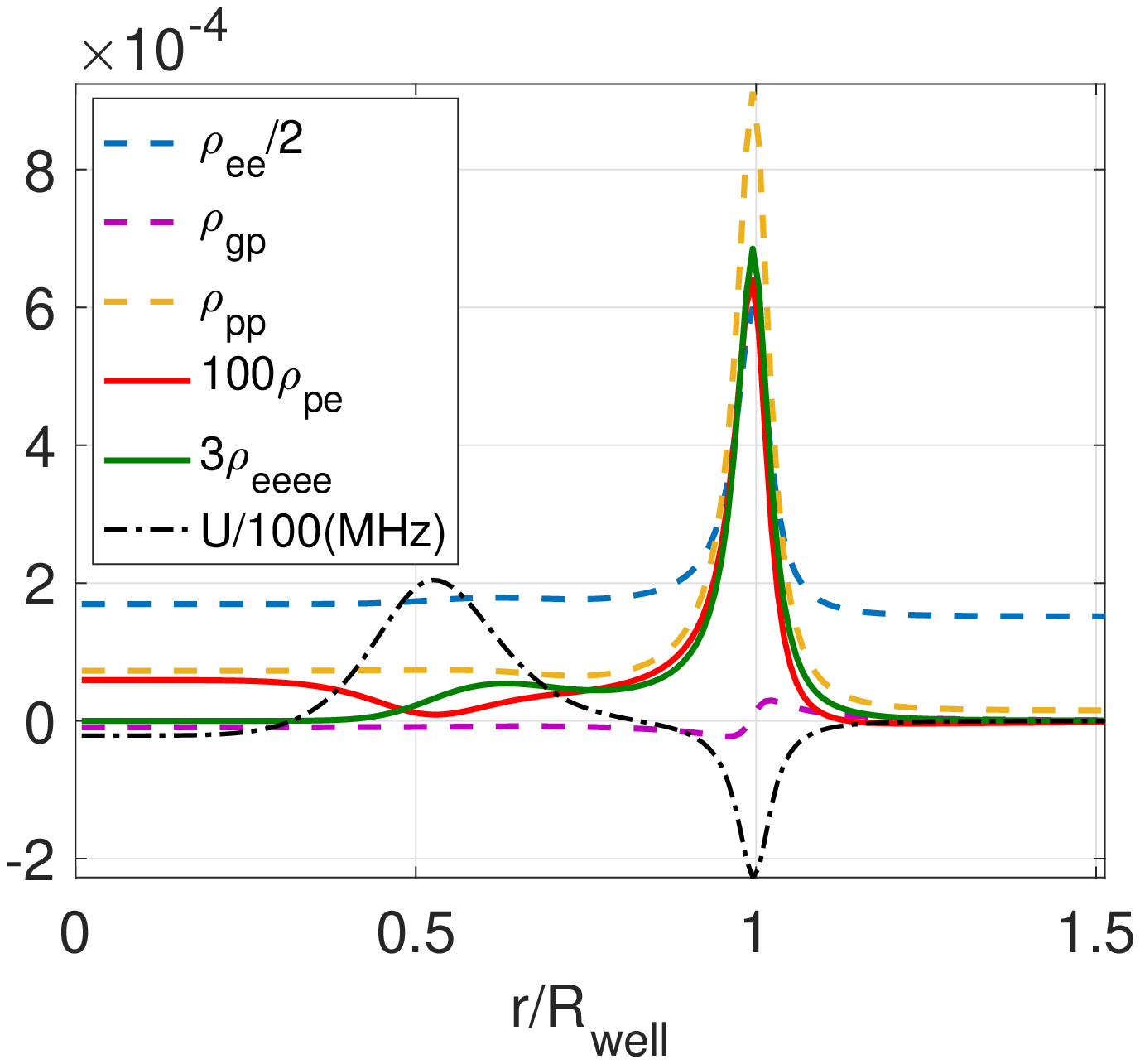}}\hfill 
\caption{ Explaining the potential profiles in Fig. \ref{Potential} based on density matrix elements. The evolution of steady state density matrix elements over inter-atomic separation are presented. Elements are scaled for comparison, see legends for the scaling factors. Dashed dotted line shows the interaction in the unit of 10kHz. (a-c) The inner-peak is generated due to the increasing of two atom Rydberg excitation, adding an effective interaction of $\frac{C_6}{r^6} \rho_{ee,ee}$. While the probability of two atom excitation $\rho_{ee,ee}$ over inner-peak  is increased by blockade leakage, the total population of $\rho_{ee}$ and $\rho_{pp}$ are not affected. Therefore, loss rate is preserved over the inner-peak. The reason that the peaks of $\rho_{ee,ee}$ and U do not match is the reduction of $\frac{\Omega_2}{2} \rho_{pe+}$ interaction term.
(c) The outer-peak is  strongly affected by the increase of intermediate population leading to $-\Delta \rho_{pp}$ interaction term. Increasing both intermediate and Rydberg population results to enhanced loss rate over the outer-peak.
 Laser parameters in Fig. [a,b,c] are $\Delta/2\pi=[10,10,3]$MHz, $\Omega_2/2\Delta=[1,0.7,2.5]$, n=100, $\Gamma_c=50$Hz, lasers are locked with $\gamma_{1,2}=100\gamma_p$.}\label{DensityMatrix}
\end{figure}

 Laser de-phasing induced sign-change was discussed in Fig. 2 for the case with fixed loss rate $\Gamma=100$Hz. 
  The effective light-shift is described by Eqs. 1,2 where 
\begin{equation}
\bar{U}_{an}(r)=2[\frac{\Omega_1}{2}\rho_{gp_{+}}+\frac{\Omega_2}{2}\rho_{pe_{+}}-\Delta \rho_{pp}] +V(r) \rho_{ee,ee}
\label{EqInteraction}
\end{equation}
where $\rho_{pe_{+}}=\rho_{pe}+\rho_{ep}$. The evolution of steady state density matrix elements are plotted in Fig. \ref{SignChangeRho}b  as a function of laser de-phasing $\gamma_2$. 
The soft-core is generated by blockade effect, resulting to negligible  $\rho_{ee,ee}$ within the radius $R_c$, see Fig. \ref{SignChangeRho}b. 
The following  investigates   one atom density matrix elements within the soft-core, having the interaction dependent value of $\rho_{pp}$ fixed numerically (by adjusting $\Omega_1$) to keep the loss constant.
Considering the assumptions of $\Omega_1/\Omega_2\ll1$ and $\Omega_2/2\Delta=1$ the main defining terms in Eq. \ref{EqInteraction} are $\rho_{pe}$ and $\rho_{pp}$.  When de-coherence terms are smaller than EIT bandwidth, the relation of $\rho_{pe}$ and $\rho_{pp}$ in a single dressed atom is given by
\begin{equation}
\rho_{pe_{+}}=\frac{4\Delta}{\Omega_2}\frac{ \gamma_p}{\gamma_2+\gamma_p}\rho_{pp}.
\label{Eqrho23simple}
\end{equation}
Using the numerical value of $\rho_{pp}$ in the presence of interaction, the above formula has a similar trend comparing to the numeric results of Fig. \ref{SignChangeRho}b. Looking at the Eq.\ref{EqInteraction} for the case of $\Omega_2/2\Delta=1$, the collective light shift has the form of $\bar{U}_{c}=\Delta \rho_{pp}\frac{\gamma_p-\gamma_2}{\gamma_p+\gamma_2}$  that starts from $\Delta \rho_{pp}$ then goes to zero at $\gamma_2=\gamma_p$ and then goes to $-\Delta \rho_{pp}$, see Fig. 2a. 
The reason for sign-change is the change in $\rho_{pe_{+}}$ that drops over the range of $0.01\gamma_p<\gamma_2< 100\gamma_p$ with the half value at $\gamma_2=\gamma_p$, see Fig. \ref{SignChangeRho}b.

At large dephasing rates $\gamma_{2}$ the EIT breaks down. This can be seen by looking at the simplified version of $\rho_{ge_{+}}$ in one atom dressing
\begin{equation}
\rho_{ge_{+}}=\frac{-\frac{2 \Omega_2}{\Omega_1} \gamma_p \rho_{pp}+\frac{2 \Omega_1}{\Omega_2} \gamma_r \rho_{ee}}{\gamma_{\text{lock}}+\gamma_r}\approx \frac{-\frac{2 \Omega_2}{\Omega_1} \gamma_p \rho_{pp}}{\gamma_{\text{lock}}+\gamma_r}, 
\label{Eqrho13}
\end{equation}
Without laser locking ($\gamma_{\text{lock}}=\gamma_1+\gamma_2$), the EIT  gets disturbed and $\rho_{ge}$ goes to zero when $\gamma_2/\gamma_p \gg \frac{\Omega_2}{\Omega_1}$. 
By locking the laser, $\rho_{ge_{+}}$ would preserve for a wider $\gamma_2$ range.

\subsection{Phase dependent interaction of solitons}
\begin{figure}
\centering
      \subfloat{%
    \includegraphics[width=.24\textwidth]{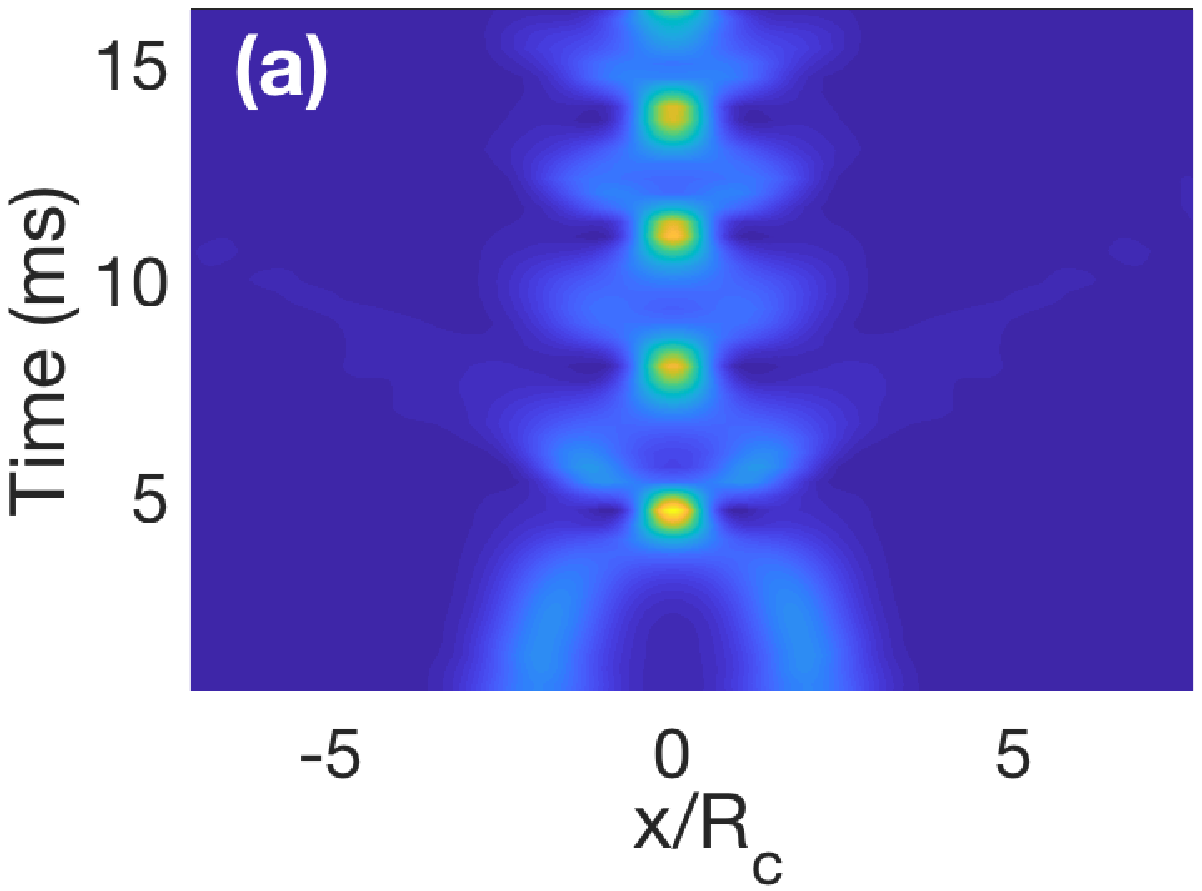}}\hfill  
      \subfloat{%
    \includegraphics[width=.24\textwidth]{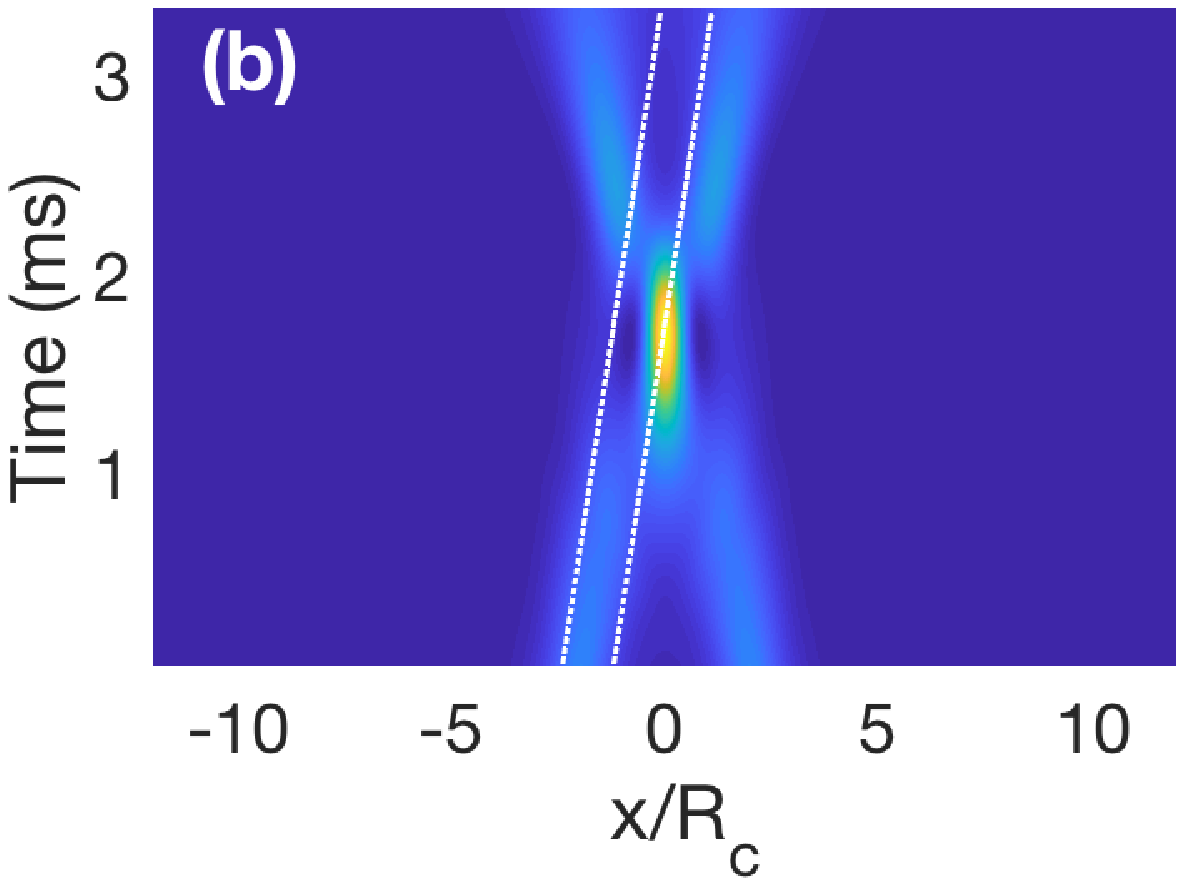}}\hfill 
          \subfloat{%
    \includegraphics[width=.24\textwidth]{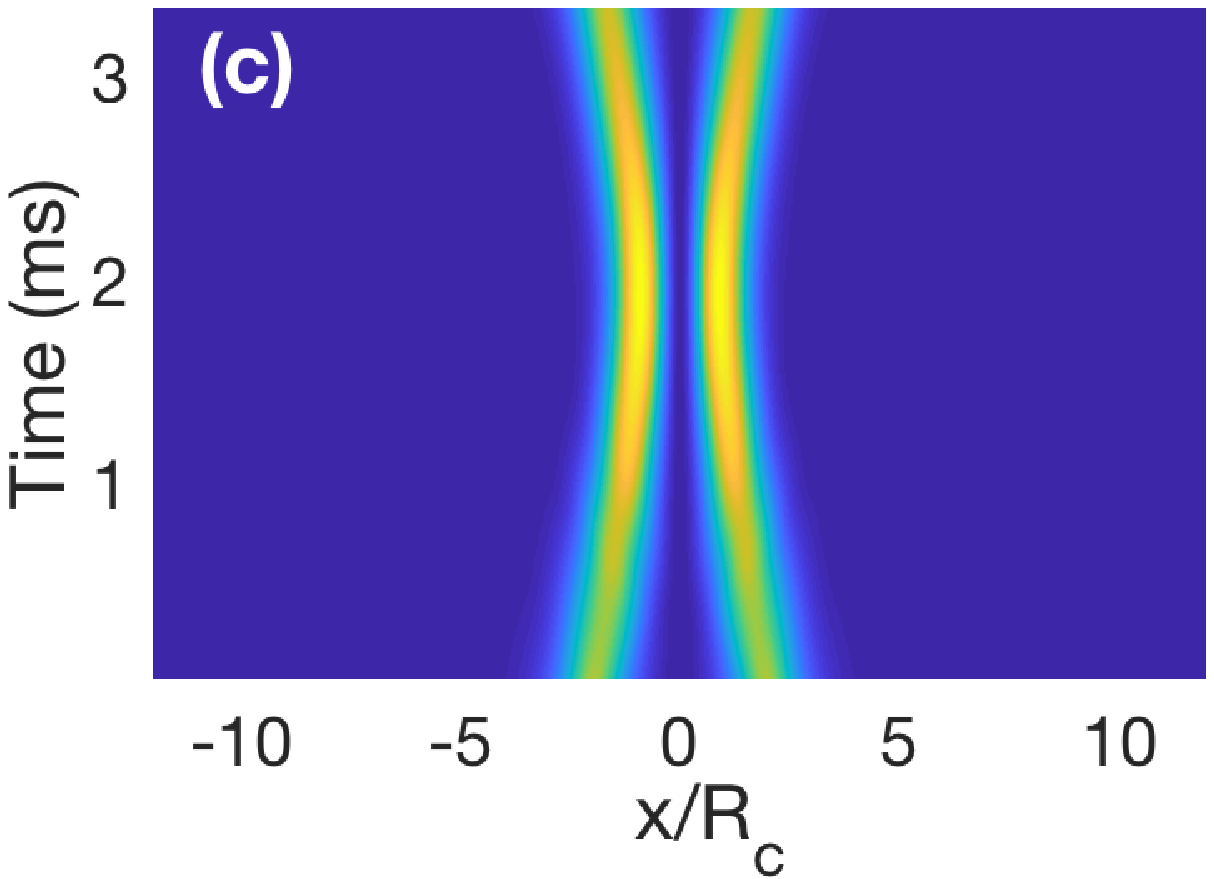}}\hfill  
      \subfloat{%
    \includegraphics[width=.24\textwidth]{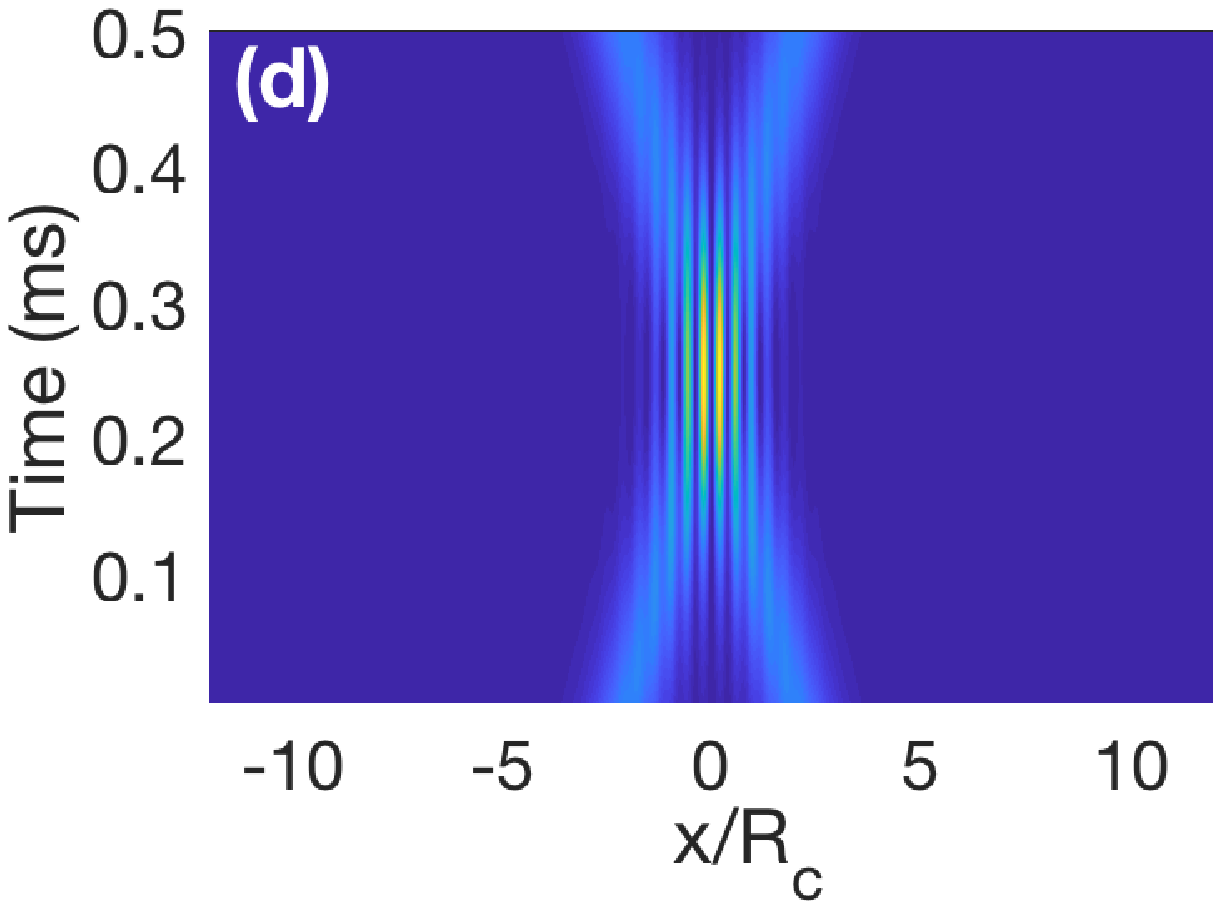}}\hfill 
          \subfloat{%
    \includegraphics[width=.24\textwidth]{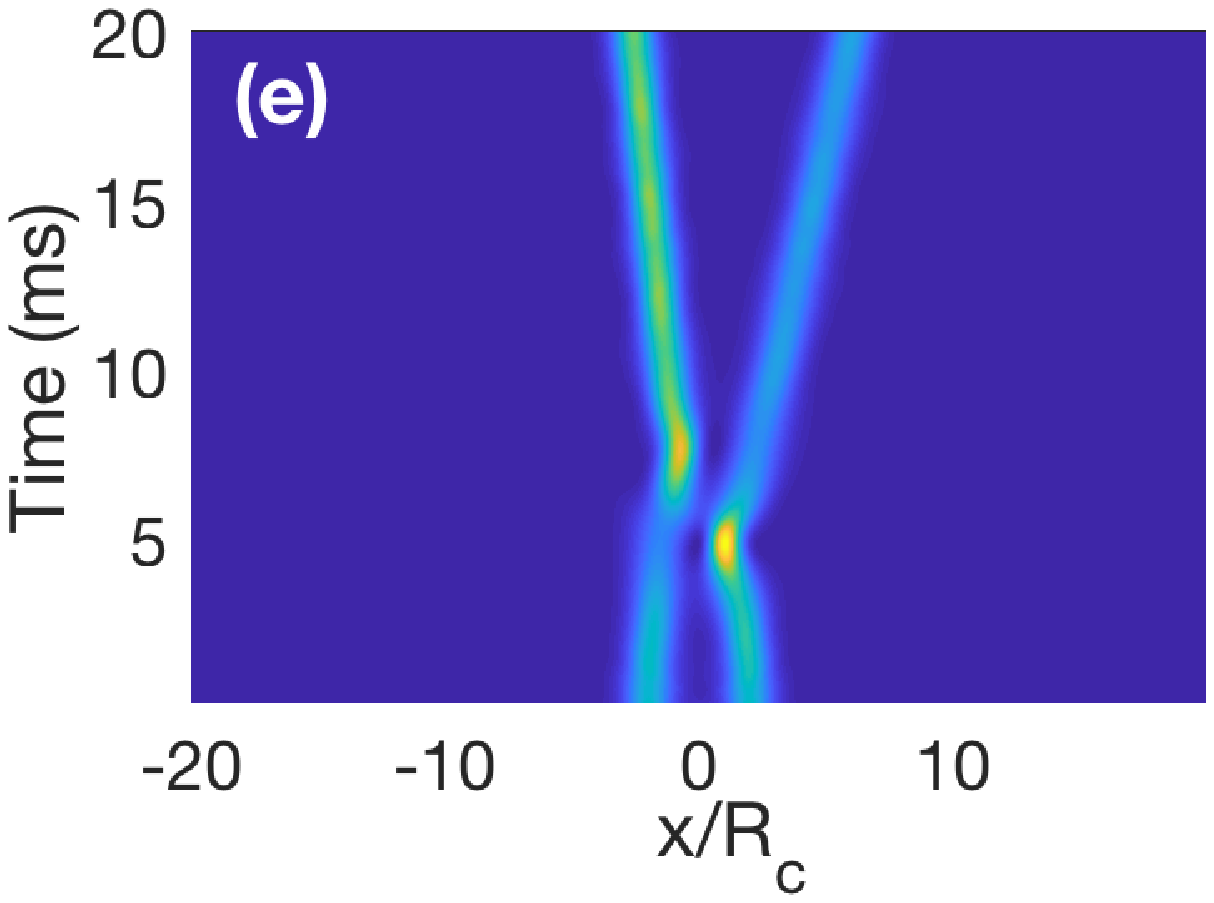}}\hfill   
\caption{Collision of bright solitons made by soft-core potential. The phase differences in first to third rows are 0, $\pi$ and $\pi/4$ respectively. (a) Soliton fusion happens between in-phase solitons.  (b) If the kinetic energy is larger than the bound energy of fused solitons, they go through each other.  While passing each other they accelerate and decelerate which results to a time advance (see dotted lines), but they will keep going with previous speed after the collision. (c) When they are out of phase, they would repel each other. (d) Above critical velocity, they would pass each other generating interference patterns. (e) Having imaginary parts in the wave-function (i.e. the phase is not exactly k$\pi$), results to mass transfer and complicated dynamics. This is the main disadvantage of using phase dependent soliton interaction.}\label{SolitonCollisionSoft}
\end{figure}

Phase dependent inter-solitonic interaction has been observed between  solitons formed by attractive local nonlinearities in BEC \cite{Ngu14}.
Here the phase dependent interaction is studied between Rydberg solitons that are formed by conventional plateau type interactions e.g. $U1$ in Fig. \ref{Potential}b. 
Fig. \ref{SolitonCollisionSoft} simulates the collision of two bright solitons under soft-core potential $U=\frac{U_C}{1+(\frac{x-x'}{R_c})^6}$ with two solitons' initial state of   $\psi_0(x)=A\sech(\sigma_r(x-x_0)) \text{e}^{\text{i}( v_r x+ \phi_r)}+A\sech(\sigma_l(x+x_0)) \text{e}^{\text{i}( v_l x+ \phi_l)}$ that is normalized to particle number  $\int |\psi_0|^2 \text{d}x=N$.   
The interaction in this case is based on coherent interference of wave-function. 
When two solitons are  out of phase $\phi_l-\phi_r=\pi$, the destructive interference upon solitons' tail over-lap, reduces the combined probability amplitude.  As a result the potential energy would increase under the attractive non-linear soft-core interaction. This would result to quantum repulsion as depicted in  Fig. \ref{SolitonCollisionSoft}c,d. This repulsion comes from the quantum coherence and does not have classical counter parts. Fast collisions with relative speeds exceeding the critical velocity, make solitons to pass through each other and generate interference pattern. Corresponding fringes' width are defined by solitons' relative velocity.
 On the other hand, when colliding solitons are in phase, their amplitude would have constructive interference. This would reduce total energy that depends on the atomic density and therefore results to attraction as shown in Fig.  \ref{SolitonCollisionSoft}a,b. Depending on collision velocity, solitons would either fuse (Fig. \ref{SolitonCollisionSoft}a) or go through each other (Fig. \ref{SolitonCollisionSoft}b) with a time advanced feature. Critical velocity is defined by the comparison of the kinetic energy and the bound energy of fused solitons.
Relative phases other than $\Delta \phi=0,\pi$ result to complicated dynamics including mass transfer, see Fig. \ref{SolitonCollisionSoft}e. While this type of interaction is vulnerable to phase fluctuations, it could be used for making  photonic and matter phase beam splitters. 

\end{document}